\documentclass[%
preprint, 
longbibliography, 
amsmath,amssymb,
aps,
prb,
superscriptaddress,
showpacs, showkeys]{revtex4-1}

\let\oldaddcontentsline\addcontentsline
\newcommand{\stoptocentries}{\renewcommand{\addcontentsline}[3]{}}
\newcommand{\starttocentries}{\let\addcontentsline\oldaddcontentsline}

\newenvironment{changemargin}[2]{
\begin{list}{}{%
\setlength{\topsep}{0pt}%
\setlength{\leftmargin}{#1}%
\setlength{\rightmargin}{#2}%
\setlength{\listparindent}{\parindent}%
\setlength{\itemindent}{\parindent}%
\setlength{\parsep}{\parskip}%
}%
\item[]}{\end{list}}

\usepackage{amsmath,environ}
\NewEnviron{equationsenviron}{
\begin{equation}\begin{split}
  \BODY
\end{split}\end{equation}
}

\usepackage{bm,array}
\usepackage{booktabs} 
\usepackage{multirow} 
\usepackage{graphicx}
\usepackage[table,xcdraw]{xcolor} 
\usepackage{float} 
\usepackage{textcomp}
\usepackage{afterpage} 
\usepackage{verbatim} 
\usepackage{color} 
\usepackage{soul}
\usepackage{amsmath,amssymb}
\newcommand{\angstrom}{\textup{\AA}} 
\usepackage{graphicx}
\usepackage{dcolumn}
\usepackage{bm}
\usepackage{hyperref}
\hypersetup{bookmarksnumbered, pdfpagemode=UseOutlines, 
colorlinks=true, citecolor=blue, filecolor=blue, linkcolor=blue, urlcolor=blue}

\usepackage{breqn}
\usepackage{bm}
\usepackage{xr}

\usepackage{booktabs} 
\usepackage{multirow}
\usepackage[table,xcdraw]{xcolor} 
\usepackage{float} 
\usepackage{siunitx} 
\usepackage{textcomp}
\usepackage{afterpage} 
\usepackage{verbatim} 
\usepackage{color} 
\usepackage{soul}
\usepackage{amsmath,amssymb}
\usepackage{graphicx}
\usepackage{dcolumn}
\usepackage{bm}
\usepackage{hyperref}
\usepackage{xr}
\hypersetup{bookmarksnumbered, pdfpagemode=UseOutlines, 
colorlinks=true, citecolor=blue, filecolor=blue, linkcolor=blue, urlcolor=blue}

\newcommand{\beginsupplement}{%
        \setcounter{table}{0}
        \renewcommand{\thetable}{S\arabic{table}}%
        \setcounter{figure}{0}
        \renewcommand{\thefigure}{S\arabic{figure}}
        \setcounter{equation}{0}
        \renewcommand{\theequation}{S\arabic{equation}}%
     }
     
     \makeatletter
\let\cat@comma@active\@empty
\makeatother

\begin{document}
\newcolumntype{C}{>{\centering\arraybackslash}p{2em}}

\title{Bias induced up to 100\% spin-injection and detection polarizations in ferromagnet/bilayer-hBN/graphene/hBN heterostructures}




\author{M. Gurram}
\thanks{corresponding author}
\email{m.gurram@rug.nl}
\affiliation{Physics of Nanodevices, Zernike Institute for Advanced Materials, University of Groningen, The Netherlands}
\author{S. Omar}
\thanks{corresponding author}
\email{s.omar@rug.nl}
\affiliation{Physics of Nanodevices, Zernike Institute for Advanced Materials, University of Groningen, The Netherlands}
\author{B.J. van Wees}
\affiliation{Physics of Nanodevices, Zernike Institute for Advanced Materials, University of Groningen, The Netherlands}%

\date{\today}
\begin{abstract}
We study spin transport in a fully hBN encapsulated monolayer-graphene van der Waals (vdW) heterostructure, at room temperature. A top-layer of bilayer-hBN is used as a tunnel barrier for spin-injection and detection in graphene with ferromagnetic cobalt electrodes. We report surprisingly large and bias induced (differential) spin-injection~(detection) polarizations up to 50$\%$~(135$\%$) at a positive voltage bias of +0.6 V, as well as sign inverted polarizations up to -70$\%$~(-60$\%$) at a reverse bias of -0.4~V. This demonstrates the potential of bilayer-hBN tunnel barriers for practical graphene spintronics applications. With such enhanced spin-injection and detection polarizations, we report a record two-terminal (inverted) spin-valve signals up to 800 $\Omega$ with a magnetoresistance ratio of 2.7$\%$, and we achieve spin accumulations up to 4.1 meV. We propose how these numbers can be increased further, for future technologically relevant graphene based spintronic devices.

\stoptocentries
\begin{description}
\item[PACS numbers]
 \verb+85.75.-d+, \verb+73.22.Pr+, \verb+75.76.j+, \verb+73.40.Gk+
\end{description}
\end{abstract}

\maketitle

Recent progress in the exploration of various two-dimensional materials has led to special attention for van der Waals (vdW) heterostructures for advanced graphene spintronics devices. For graphene spin-valve devices, an effective injection and detection of spin-polarized currents with a ferromagnetic (FM) metal via efficient tunnel barriers is crucial\cite{92_Roche2015_FlaghsiP_rev,89_DattaDas1990_APL}. The promising nature of crystalline hBN layers as pin-hole free tunnel barriers\cite{65_Britnell2012_Sci_GrhBNGr} for spin injection into graphene\cite{35_Gurram2016_PRB,32_Yamaguchi2013_APE,36_Singh2016_APL,33_Kamalakar2014_SciRep,34_Fu2014_JAP} has been recently demonstrated. However, due to the relatively low interface resistance-area product of monolayer-hBN barriers, there is a need to use a higher number of hBN layers for non-invasive spin injection and detection\cite{50_Thomas2012_PRB}. Theoretically, large spin-injection polarizations have been predicted in FM/hBN/graphene systems as a function of bias with increasing number of hBN layers\cite{17_Wu2014_PRA}.

Kamalakar \textit{et al.}\cite{19_Kamalakar2016_SciRep} reported an inversion of the spin-injection polarization for different thicknesses of chemical vapour deposited (CVD)-hBN tunnel barriers, as well as an asymmetric bias dependence of the polarization using multilayer CVD-hBN/FM tunnel contacts. The observed behaviour was attributed to spin-filtering processes across the multilayer-hBN/FM tunnel contacts. 

Here we explicitly show the unique role of current/voltage bias for spin-injection and detection through bilayer(2L)-hBN tunnel barriers with ferromagnetic cobalt (Co) electrodes. Application of a bias across the FM/hBN/graphene tunneling contacts a) allows to widen the energy window up to $\sim$ 1~eV for an additional spin polarized states in the FM and graphene to participate in the tunneling spin-injection and detection processes, b) induces a large electric-field between the FM and graphene which can modify the tunneling processes, c) provides electrostatic gating for the graphene which could change the carrier density between electrons and holes, and d) is predicted to induce magnetic proximity exchange splitting in graphene of up to 20 meV\cite*{6_Zollner2016_PRB, 16_Lazi2016_RapCom}.


We show that bilayer-hBN tunnel barriers are unique for spin-injection and detection in graphene, with (differential) polarizations unexpectedly reaching values close to $\pm$ 100\% as a function of the applied DC bias at room temperature. Furthermore, we demonstrate a two-terminal (inverted) spin-valve with a record magnitude of the spin signal reaching 800 $\Omega$ with magnetoresistance ratio of 2.7$\%$.
\stoptocentries  
\section{Four-terminal non-local spin transport}
We study the spin transport in fully hBN encapsulated graphene, prepared via dry pick-up and transfer method\cite{44_Zomer2014_APL} to obtain clean and polymer free graphene-hBN interfaces\cite{35_Gurram2016_PRB} (see Methods for device fabrication details). We use a four-terminal non-local measurement geometry to separate the spin current path from the charge current path (Fig.~\ref{fig:Figure_1}a). An AC current ($i$) is applied between two Co/2L-hBN/graphene contacts to inject a spin-polarized current in graphene. The injected spin accumulation in graphene diffuses and is detected non-locally ($v$) between the detector contacts using a low frequency (f=10-20 Hz) lock-in technique. For the spin-valve measurements, the magnetization of all the contacts is first aligned by applying a magnetic field $B_{y}$ along their easy axes. Then $B_{y}$ is swept in the opposite direction. The magnetization reversal of each electrode at their respective coercive fields appears as an abrupt change in the non-local differential resistance $R_{nl}$ (= $v/i$). Along with a fixed amplitude $i$ of 1-3 $\mu$A, we source a DC current ($I_{in}$) to vary the bias applied across the injector contacts. In this way, we can obtain the differential spin-injection polarization of a contact, defined as $p_{in}=\frac{i_{\text{s}}}{i}=\frac{dI_{\text{s}}}{dI}$ where $I_{\text{s}}$ ($i_{s}$) are the DC (AC) spin currents, and study in detail how $p_{in}$ of the contacts depends on the applied bias. We observe that the magnitude of the differential spin signal $\Delta R_{nl}$ at a fixed AC injection current increases with the DC bias applied across the injector (Fig.~\ref{fig:Figure_2}a,~\ref{fig:Figure_2}c). Moreover, a continuous change in the magnitude of $\Delta R_{nl}$ between -4.5 $\Omega$ and 2.5 $\Omega$ as a function of DC current bias across the injector, and its sign reversal close to zero bias can be clearly observed (Fig.~\ref{fig:Figure_3}a). A similar behaviour is also observed for different injection contacts.

In Hanle spin-precession measurements, where the magnetic field $B_{z}$ is swept perpendicular to the plane of spin injection, the injected spins precess around the applied field and dephase while diffusing towards the detectors. We obtain the spin transport parameters such as spin-relaxation time $\tau_{s}$ and spin-diffusion constant $D_{s}$ by fitting the non-local Hanle signal $\Delta R_{nl}(B_{z})$ with the stationary solutions to the steady state Bloch equation in the diffusion regime; $D_{s}\bigtriangledown ^{2}\overrightarrow{\mu_{s}} - \overrightarrow{\mu_{s}}/\tau_{s} + \gamma \overrightarrow{B_{z}}\times \overrightarrow{\mu_{s}}= 0$. Here, the net spin accumulation $\mu_{s}$ is the splitting of spin chemical potentials spin-up $\mu_{\uparrow}$ and spin-down $\mu_{\downarrow}$, i.e., ($\mu_{\uparrow}-\mu_{\downarrow}$)/2 and $\gamma$ is the gyromagnetic ratio.
In order to obtain reliable fitting parameters, we probe the Hanle signals for a long spin transport channel of length $L$ = 6.5 $\mu$m. We measure the Hanle signals for different DC current bias and see that the signals cross zero at the same values of $B_{z}$ (Fig.~\ref{fig:Figure_2}d). This implies that $\tau_{s}$ and $D_{s}$ in the channel are not affected by the bias across the injector. We obtain $\tau_{s} \sim $ 0.9 ns, $D_{s} \sim $ 0.04 m$^2$/s, and $\lambda_{s} \sim $ 5.8 $\mu$m. We estimate the carrier density $n \simeq 5 \times 10^{12}$ cm$^{-2}$ from the Einstein relation by assuming $D_{s}$ = $D_{c}$\cite{94_Weber2005_Nat_DcDs}, where $D_{c}$ is the charge diffusion constant (see supplementary information section-VIII (SI-VIII)).
\section{Spin-injection polarization}
Since $\lambda_{s}$ does not change due to the bias applied between the injector contacts, the bias dependence of the non-local differential spin signal $\Delta R_{nl}$ in Fig.~\ref{fig:Figure_2} and Fig.~\ref{fig:Figure_3}a is due to the change in spin-injection polarization. From $\Delta R_{nl}$ in Fig.~\ref{fig:Figure_3}a, we can obtain the differential spin-injection polarization of the injector contact 8, $p_{in}^{8}$ from\citep{23_Tombros2007_Nat} 
 \begin{equation}
   \Delta R_{\text{nl}}^{8-9} = \frac{R_{\text{sq}}\lambda_{s}}{2W} \left[ p_{in}^{8} p_{d}^{9}e^{\frac{-L}{\lambda_{\text{s}}}} \right] , 
 \label{dRnl_89_main}
  \end{equation}
using a known unbiased detection polarization of detector 9, $p_{d}^{9}$ (see SI-III for the analysis and calculation of $p_{d}^{9}$), the length between contacts 8 and 9, $L_{8-9}=1~\mu$m, the square resistance $R_{sq} \sim$ 400 $\Omega$, and the width W = 3 $\mu$m of graphene. The non-local spin signal as a function of bias due to the spin injection through 8 is obtained from $\Delta R_{nl}^{\text{8-9}}(I_{in})$ = $(R_{nl}^{\text{$\uparrow$$\uparrow$$\uparrow$}}(I_{in}) - R_{nl}^{\text{$\uparrow$$\downarrow$$\uparrow$}}(I_{in}))/2 $, where $R_{nl}^{\text{$\uparrow$$\uparrow$$\uparrow$}}(I_{in})$ is the non-local signal measured as a function of $I_{in}$ when the magnetization of contacts 7, 8, and 9 are aligned in $\uparrow$ , $\uparrow$ , and $\uparrow$ configuration, respectively. We find that $p_{in}^{8}$ changes from -1.2\% at zero bias to +40\% at +25 $\mu$A and -70\% at -25 $\mu$A (Fig.~\ref{fig:Figure_3}b). It shows a sign inversion which occurs close to zero bias. The absolute sign of $p$ cannot be obtained from the spin transport measurements and we define it to be positive for the majority of the unbiased contacts (SI-III). 

The observed behaviour of the (differential) polarization is dramatically different from what has been observed so far for spin-injection in graphene, or in any other non-magnetic material. For spin-injection/detection with conventional ferromagnetic tunnel contacts, the polarization does not change its sign close to zero bias. It can be modified at high bias\cite{9_Valenzuela5_PRL}. However, in our case we start with a very low polarization at zero bias which can be enhanced dramatically in positive and negative directions.

The above analysis is repeated for other bilayer-hBN tunnel barrier contacts with different interface resistances. Figure~\ref{fig:Figure_4}a shows $p_{in}$ for four contacts plotted as a function of the voltage bias obtained from the respective $\Delta R_{nl}(I_{in})$. All contacts show similar behaviour, where the magnitude of $p_{in}$ increases with bias and changes sign close to zero bias. For the same range of the applied voltage bias, contacts with either 1L-hBN or TiO$_{2}$ tunnel barriers do not show a significant change in the spin polarization (SI-VI and SI-X). This behaviour implies that the observed tunneling spin-injection polarization as a function of the bias is unique to bilayer-hBN tunneling contacts.

\section{Spin-detection polarization}
\begin{figure}[H] 
\centering
 \includegraphics[width=\columnwidth,trim= 0in 0in 0in 0in,clip]{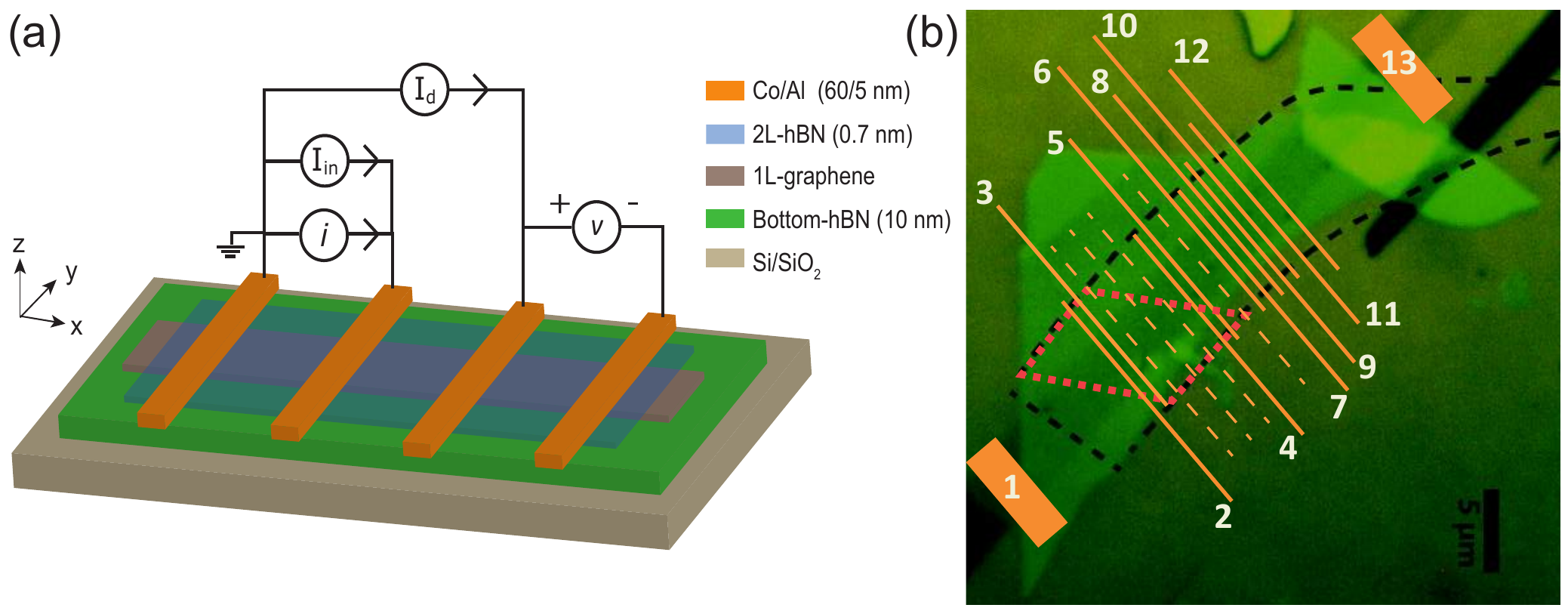}
 \caption{\label{fig:Figure_1} \textbf{Device layout and measurement scheme}. (\textbf{a}) A layer by layer schematic of the vdW heterostructure of the 2L-hBN/graphene/thick-hBN stack with FM cobalt electrodes. A measurement scheme is shown for the non-local spin transport measurements with a DC current bias $I_{in}$ and AC current $i$, applied across the injector contacts and a non-local differential (AC) spin signal $v$ is measured using a lock-in detection technique. A DC current bias $I_{d}$ can also be applied in order to bias the detector contact.
(\textbf{b}) An optical microscopic picture of the vdW heterostructure. The black-dashed line outlines the hBN tunnel barrier flake. The red-dashed line outlines the monolayer region of the hBN tunnel barrier flake (see SI-I for the optical microscopic picture of the tunnel barrier). A schematic of the deposited cobalt electrodes is shown as orange bars and the Co/hBN/graphene contacts are denoted by numbers 1, 2,.., and 13. The orange-dashed lines represent the unused contacts. Cobalt electrodes from 2 to 5 are either fully or partially deposited on top of the monolayer region of the tunnel barrier flake while the electrodes from 6 to 12 are exclusively deposited on the bilayer region. The width of the cobalt electrodes (2 to 12) is varied between 0.15 and 0.4 $\mu$m.}
\end{figure}

\begin{figure}[H] 
\setlength{\abovecaptionskip}{-6pt}
\centering
 \includegraphics[width=0.9\columnwidth,trim= 0in 0in 0in 0in,clip]{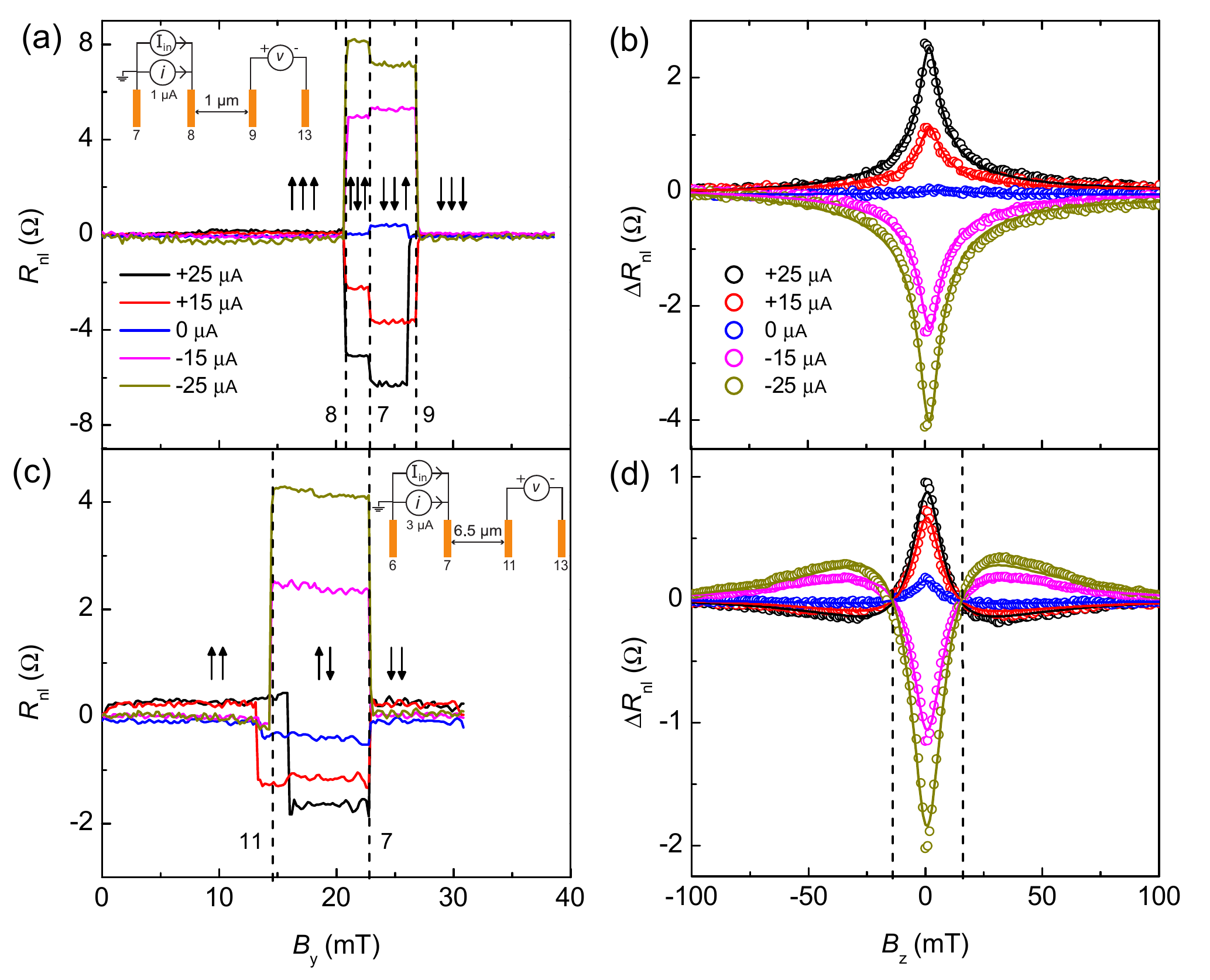}
\vspace*{-1mm} %
 \caption{\label{fig:Figure_2} \textbf{Non-local spin-valve and Hanle measurements at different DC bias across the injector.} (\textbf{a},\textbf{c}): Non-local differential spin-valve signal $R_{nl}$ (= $v$/$i$) as a function of the magnetic field $B_{y}$ applied along the easy axes of the Co electrodes, for a short (L = 1 $\mu$m) (\textbf{a}), and a long (L = 6.5 $\mu$m) (\textbf{c}) spin transport channel. An offset at zero field is subtracted from each curve for a clear representation of the data. The vertical dashed lines correspond to the switching of the electrodes at their respective coercive fields. The switch of the outer detector 13 is not detectable as it is located far ($>$ 2$\lambda_{s}$) from the nearest injector. The legend shows the applied injection DC current bias $I_{in}$ values. The up($\uparrow$) and down($\downarrow$) arrows represent the relative orientation of the electrode magnetizations. The three arrows in (\textbf{a}) correspond to the contacts 7, 8, and 9, and the two arrows in (\textbf{c}) correspond to the contacts 7 and 11, from left to right. The insets show the measurements schematics, injection AC current ($i$) and the DC current bias ($I_{in}$), the respective contacts used for the spin current injection, and non-local differential voltage ($v$) detection. The differential spin signal in \textbf{a} due to spin injection through 8 is $\Delta R_{nl}^{\text{8-9}}$ = $(R_{nl}^{\text{$\uparrow$$\uparrow$$\uparrow$}} - R_{nl}^{\text{$\uparrow$$\downarrow$$\uparrow$}})/2$, and in \textbf{c} due to spin injection through 7 is $\Delta R_{nl}^{\text{7-11}}$ = $(R_{nl}^{\text{$\uparrow$$\uparrow$}} - R_{nl}^{\text{$\uparrow$$\downarrow$}})/2$.
(\textbf{b},\textbf{d}): Non-local (differential) Hanle signal $\Delta R_{nl}(B_{z})$ as a function of the magnetic field $B_{z}$. \textbf{b}(\textbf{d}) shows $\Delta R_{nl}$ measured for the short(long) channel, corresponding to the spin injector contact 8(7) and measured with the detector contact 9(11). The measured data is represented in circles and the solid lines represent the fits to the data. Hanle signals in \textbf{b} at different injection bias values $\Delta R_{nl}^{\text{8-9}}(B_{z})$ = $(R_{nl}^{\text{$\uparrow$$\uparrow$$\uparrow$}}(B_{z}) - R_{nl}^{\text{$\uparrow$$\downarrow$$\uparrow$}}(B_{z}))/2 $. The two vertical dashed lines in \textbf{d} correspond to the fields where the Hanle signals cross zero.}
\end{figure}

\begin{figure}[H]
\centering
 \includegraphics[width=\columnwidth,trim= 0in 0in 0in 0in,clip]{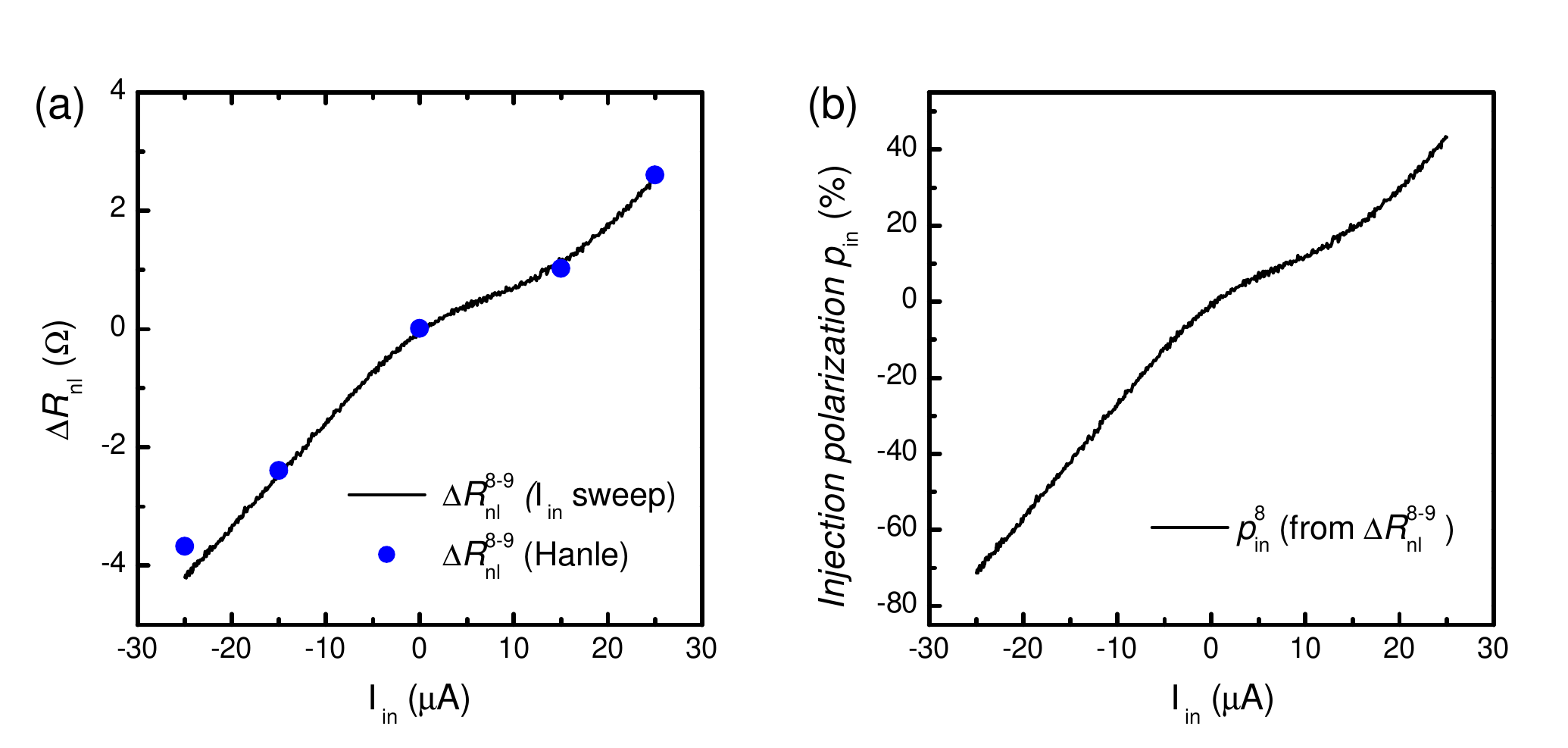}
 \caption{\label{fig:Figure_3} \textbf{Bias enhanced non-local differential spin signal and large differential spin-injection polarization at room temperature.} (\textbf{a}) Non-local spin signal $\Delta R_{nl}^{\text{8-9}}$($I_{in}$) corresponding to the spin current injected through contact 8 and detected via contact 9, as a function of the DC current bias ($I_{in}$) applied across the injector. The solid line represents the spin signal $\Delta R_{nl}^{\text{8-9}}$($I_{in}$) for a continuous sweeping of the $I_{in}$ bias, while the dots are extracted from the Hanle signals $\Delta R_{nl}^{\text{8-9}}$($B_{z}$) at $B_{z}$=0, measured at different bias (from Fig.~\ref{fig:Figure_2}b).
(\textbf{b}) Differential spin-injection polarization of the injector contact 8, $p_{in}^{\text{8}}$ as a function of $I_{in}$, calculated from the $\Delta R_{nl}^{\text{8-9}}$($I_{in}$) (Eq.~\ref{ac spin signal main}) data plotted in \textbf{a}.}
\end{figure}

\begin{figure}[H]
\centering
 \includegraphics[width=\columnwidth,trim= 0in 0in 0in 0in,clip]{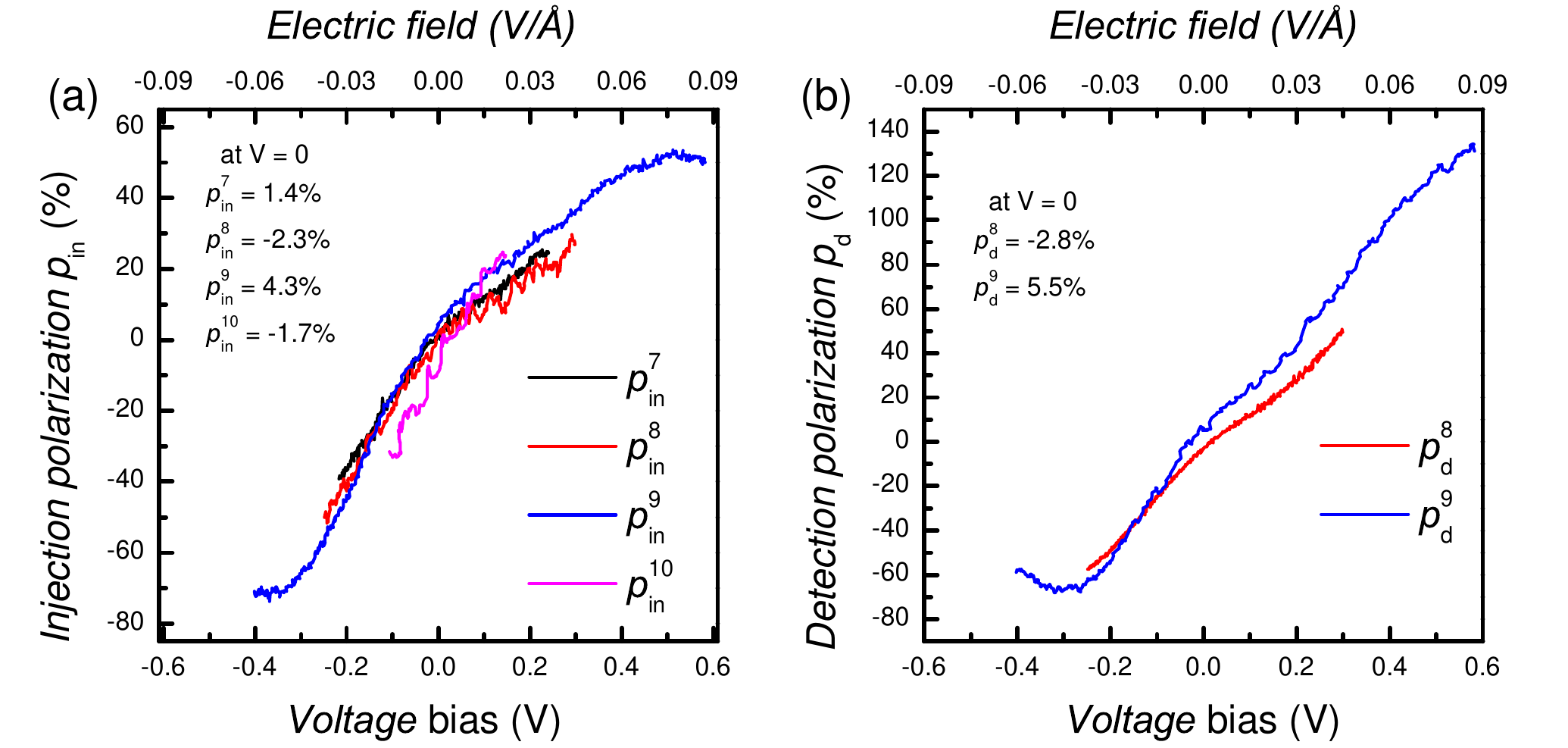}
 \caption{\label{fig:Figure_4} \textbf{Differential spin-injection ($p_{in}$) and detection ($p_{d}$) polarizations of the cobalt/bilayer-hBN/graphene contacts.} (\textbf{a}) Differential spin-injection polarization $p_{in}$ of four contacts with 2L-hBN tunnel barrier, as a function of the DC voltage bias $V$. Top axis represents the corresponding electric-field (=$V$/$t_{\text{hBN}}$, $t_{\text{hBN}}$ $\approx$ 7~$\angstrom$, the thickness of 2L-hBN barrier) induced across the Co/2L-hBN/graphene contacts. Note that the $\Delta R_{nl}$ used to calculate $p_{in}^{8}$ in Fig.~\ref{fig:Figure_3}b is obtained from a different data set.
(\textbf{b}) Differential spin-detection polarization $p_{d}$ of contacts 8 and 9 as a function of DC voltage bias $V$ applied across the detector while the injector bias is fixed at $I_{in}=$ +20 $\mu$A. The insets in \textbf{a} and \textbf{b} show $p_{in}$ and $p_{d}$ of contacts at zero bias, respectively.}
\end{figure}

We now study the effect of the bias on spin-detection. The (differential) spin detection polarization $p_{d}$ of a contact is defined as the voltage change ($\Delta V$) measured at the detector due to a change in the spin accumulation underneath ($\Delta \mu_{s}$) (see SI-II for the derivation and details), 
\begin{equation}
 p_{d} = \frac{\Delta V}{\Delta \mu_{\text{s}}/e }
 \label{detection polarization}	
\end{equation} 
where $\Delta V = i\left[\Delta R_{nl}^{in-d}(I_{d})\right]$ is measured as a function of the detector bias $I_{d}$, and $\Delta \mu_{\text{s}}/e = \frac{{i}R_{\text{sq}}\lambda_{\text{s}}}{2W} p_{\text{in}}e^{-L/\lambda_{\text{s}}}$. In a linear response regime at low bias, $p_{d}$ should resemble $p_{in}$ because of reciprocity. However in the non-linear regime at higher bias, they can be different. A comparison between Fig.~\ref{fig:Figure_4}a and \ref{fig:Figure_4}b shows that the bias dependence of $p_{i}$ and $p_{d}$ is similar. However, we find that $p_{d}$ of contact 9 can reach more than 100\% above \mbox{+0.4 V}. We note that the presence of a non-zero DC current in the graphene spin transport channel between injector and detector could modify $\lambda_{\text{s}}$ due to carrier drift, and consequently the calculated polarizations have a typical uncertainty of about 10\% (SI-IX). Although there is no fundamental reason that the biased detection polarizations $p_{d}$ cannot exceed 100\%
\footnote{As we show in SI-II, DC spin-injection polarization $P_{in}(I)$ $(=\frac{I_{\text{s}}}{I})$ is fundamentally restricted to $\pm$100\%. This requirement does not hold for the biased differential (AC) injection polarization $p_{in}(I)$ as well as the biased (differential) detection polarization $p_{d}(I)$.}, 
it could be that our observation of over 100\% polarization is due to effect of the drift which is expected to have a bigger effect on the accurate determination of $p_{d}(I)$ as compared to $p_{i}(I)$ (Fig.~\ref{fig:Figure_1}a).

Concluding, we have obtained a dramatic bias induced increase in both the differential spin-injection and detection polarizations, reaching values close to $\pm$100\% as a function of applied bias across the cobalt/bilayer-hBN/graphene contacts.



\section{Two-terminal local spin transport}
\begin{figure}[H]
\centering
 \includegraphics[width=\columnwidth,trim= 0in 0in 0in 0in,clip]{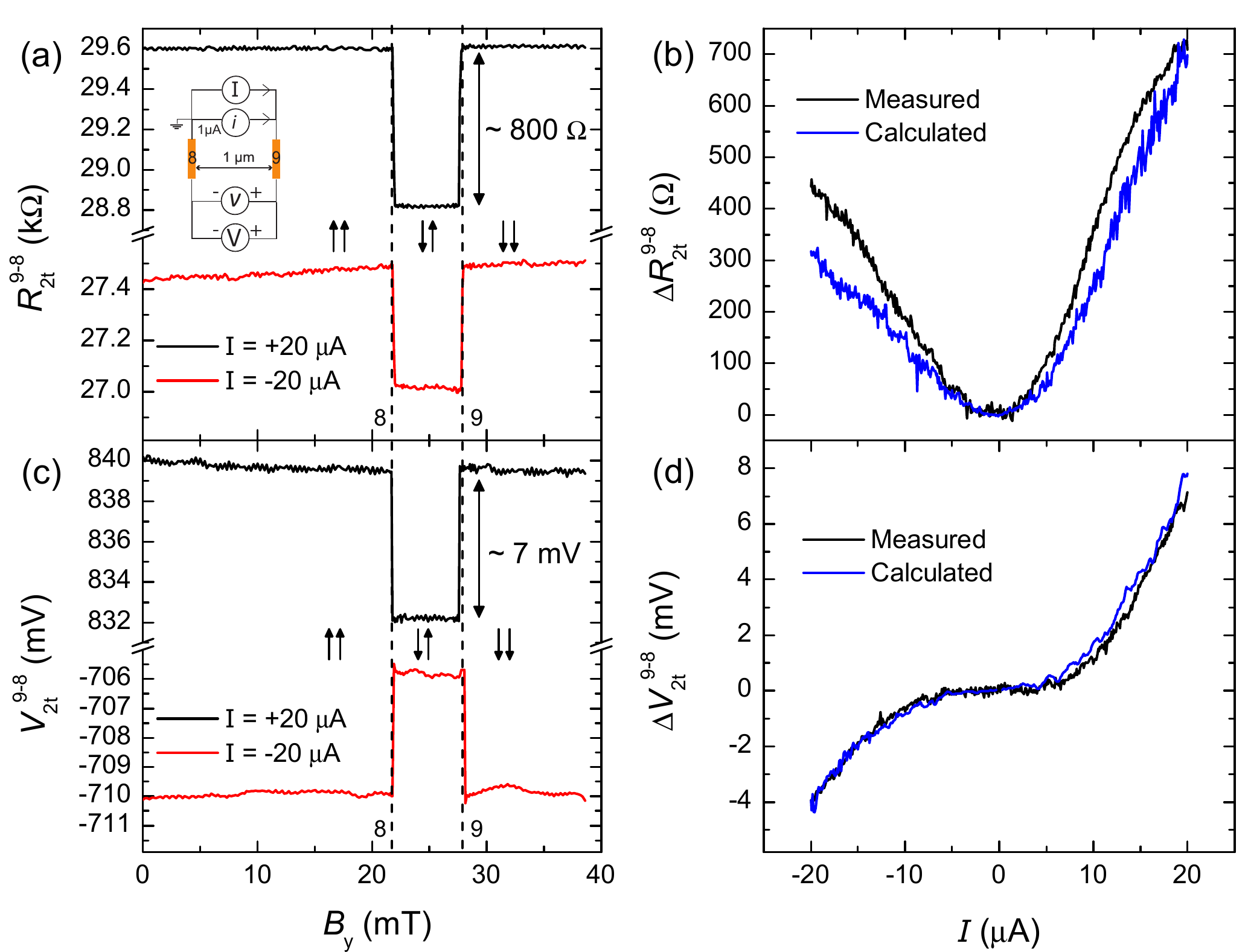}
 \caption{\label{fig:Figure_5} \textbf{Large inverted two-terminal spin-valve effect at room temperature.}
(\textbf{a}) Two-terminal differential spin-valve signal $R_{2t}$(=$v/i$) and (\textbf{c}) two-terminal DC spin-valve signal $V_{2t}$, as a function of $B_{y}$ at two different DC current bias values. The inset in \textbf{a} illustrates the two-terminal spin-valve measurement configuration. The arrows $\uparrow \uparrow$ ($\downarrow \uparrow$) represent the parallel (anti-parallel) orientation of the magnetization of contacts 8 and 9, respectively, from left to right. The vertical dashed lines represent the coercive fields of contacts 8 and 9.
(\textbf{b}) Two-terminal differential spin signal $\Delta R_{2t}(I)$ and (\textbf{d}) two-terminal DC spin signal $\Delta V_{2t}(I)$, as a function of the DC current bias $I$. The calculated two-terminal spin signals from the individual spin-injection and detection polarizations of contacts 8 and 9 are also shown in \textbf{b} and \textbf{d}.}
\end{figure}

A four-terminal non-local spin-valve scheme is ideal for proof of concept studies, but it is not suitable for practical applications where a two-terminal local geometry is technologically more relevant. In a typical two-terminal spin-valve measurement configuration, the spin signal is superimposed on a (large) spin-independent background. Since we have found that the injection and detection polarizations of the contacts can be enhanced with DC bias, the two-terminal spin signal can now be large enough to be of practical use. For the two-terminal spin-valve measurements, a current bias ($i$ + $I$) is sourced between contacts 8 and 9, and a spin signal (differential, $v$ and DC, $V$) is measured across the same pair of contacts as a function of $B_y$ (inset, Fig.~\ref{fig:Figure_5}a). Figures~\ref{fig:Figure_5}a and \ref{fig:Figure_5}c show the two-terminal differential resistance $R_{2t}$ (=$v/i$) and the two-terminal DC voltage V$_{2t}$, respectively, measured as a function of $B_y$. As a result of the two-terminal circuit, both the contacts are biased with same $I$ but with opposite polarity, resulting in opposite sign for the injection and detection polarizations. Therefore we measure an inverted two-terminal differential spin-valve signal $R_{2t}$ with minimum resistance in anti-parallel configuration. We observe a maximum magnitude of change in the two-terminal differential (DC) signal $\Delta R_{2t}$ ($\Delta V_{2t}$) of about 800 $\Omega$ (7 mV) at $I=+20$ $\mu$A, where $\Delta R_{2t}(I)$ = $R_{2t}^{\text{$\uparrow$$\uparrow$}}(I) - R_{2t}^{\text{$\downarrow$$\uparrow$}}(I)$ and $\Delta V_{2t}(I)$ = $V_{2t}^{\text{$\uparrow$$\uparrow$}}(I) - V_{2t}^{\text{$\downarrow$$\uparrow$}}(I)$ represent the difference in the two-terminal signals when the magnetization configuration of contacts 8 and 9 changes between parallel($\uparrow\uparrow$) and anti-parallel($\downarrow\uparrow$). A continuous change in $\Delta R_{2t}$ and $\Delta V_{2t}$ can be observed as a function of DC current bias (Fig.~\ref{fig:Figure_5}b and \ref{fig:Figure_5}d).


The magnetoresistance (MR) ratio of the two-terminal differential spin signal is a measure of the local spin-valve effect, and is defined as ($R_{2t}^{\text{$\downarrow$$\uparrow$}} - R_{2t}^{\text{$\uparrow$$\uparrow$}})/R_{2t}^{\text{$\uparrow$$\uparrow$}}$, where $R_{2t}^{\text{$\uparrow$$\downarrow$}}$ ($R_{2t}^{\text{$\uparrow$$\uparrow$}}$) is the two-terminal differential resistance measured in the anti-parallel (parallel) magnetization orientation of the contacts. From the spin-valve signal, we calculate the maximum MR ratio of -2.7\% at $I = +20$~$\mu$A. 

Since we have already obtained the differential spin-injection and detection polarizations of both the contacts 8 and 9 as a function of bias (Fig.~\ref{fig:Figure_4}), we can calculate the two-terminal differential spin signal from
\begin{equationsenviron}
 \triangle R_{\text{2t}}(I) = \left[ p_{in}^{9}(I_{\text{in}})p_{d}^{8}(-I_{\text{d}}) + p_{in}^{8}(-I_{\text{in}})p_{d}^{9}(I_{\text{d}}) \right] \frac{R_{\text{sq}}\lambda_{\text{s}}}{W} e^{-\frac{L}{\lambda_{\text{s}}}}
 \label{ac spin signal main} 
\end{equationsenviron} 
The calculated differential signal $\triangle R_{\text{2t}}(I)$ is plotted in Fig.~\ref{fig:Figure_5}b. A similar analysis can be done for the two-terminal DC spin signal $\triangle V_{\text{2t}}(I)$ (SI-II) and is plotted in Fig.~\ref{fig:Figure_5}d. Even though there is an uncertainty in the calculation of $p_{d}$ due to the a possible effect of carrier drift between the injector and detector, we get a close agreement between the measured and calculated signals in different (local and non-local) geometries. This confirms the accurate determination of the individual spin-injection and detection polarizations of the contacts. 

Furthermore, we can now calculate the total spin accumulation in graphene, underneath each contact in the two-terminal biased scheme, due to spin-valve effect. The results are summarized in Table~\ref{Table_spin_accumulation}. The maximum spin accumulation, beneath contact 9, due to spin-injection/extraction from contacts 8 and 9 reaches up to 4.1 meV for an applied bias of $I=20$ $\mu$A. It is noteworthy that such a large magnitude of spin accumulation in graphene at room temperature has not been reported before.

We will not speculate here on possible explanations of our fully unconventional observations. We note however that further research will require the detailed study of the injection/detection processes as a function of graphene carrier density, in particular the interaction between contact bias induced and backgate induced carrier density.
\footnote{Due to the complexity of our device, we could not realize an operating backgate.} 
Via these measurements one could also search for possible signatures of the recently proposed magnetic proximity exchange splitting in graphene with an insulator spacer, hBN\cite*{6_Zollner2016_PRB,16_Lazi2016_RapCom}.


\section{Conclusions}
In conclusion, by employing bilayer-hBN as a tunnel barrier in a fully hBN encapsulated graphene vdW heterostructure, we observe a unique sign inversion and bias induced spin-injection (detection) polarizations between 50\% (135\%) at +0.6 V and -70$\%$ (-60$\%$) at -0.4 V at room temperature. This resulted in a large change in the magnitude of the non-local differential spin signal with the applied DC bias across the Co/2L-hBN/graphene contacts and the inversion of its sign around zero bias. Such a large injection and detection polarizations of the contacts at high bias made it possible to observe the two-terminal differential and DC spin signals reaching up to 800 $\Omega$, and magnetoresistance ratio up to 2.7$\%$ even at room temperature. Moreover, we obtain a very large spin accumulation of about 4.1 meV underneath the contacts in a two-terminal spin-valve measurement.

Note that we have been conservative in biasing the contacts to prevent breakdown of the 2L-hBN barriers. By increasing the bias to the maximum theoretical limit of $\sim$ $\pm$0.8~V\cite*{87_Yoshiaki2015_NL_VbdhBN}, we expect that we can increase the polarizations even further. Also one can increase the width of the contacts by a factor of 5 to about 1 $\mu$m (yet far below $\lambda_{s}$) which will reduce the background resistance of two-terminal spin-valve signal by the same factor, and allow to apply a maximum current bias up to 100 $\mu$A\cite{79_Ingla-Aynes2016_NL}. This could result in two-terminal spin signal above 50~mV and MR ratio beyond 20\%. The corresponding change in spin accumulation could reach up to 40~meV underneath the contacts, exceeding the room temperature thermal energy ($k_{B}T$ $\sim$ 25 meV). Such high values of spin accumulation will open up an entirely new regime for studying spin transport in graphene and for applications of graphene based spintronic devices\cite*{89_DattaDas1990_APL}.

\newpage
\section{Figures}

\newpage
\vspace*{-4\baselineskip}

\begin{table}[H]
\centering
\begin{tabular}{|c|c|c|c|c|}
\hline
\multirow{2}{*}{} & \multicolumn{2}{|c|}{\textbf{\begin{tabular}[c]{@{}c@{}}$\mu_{s}$ underneath 8 \\ (meV)\end{tabular}}} & \multicolumn{2}{|c|}{\textbf{\begin{tabular}[c]{@{}c@{}}$\mu_{s}$ underneath 9 \\ (meV)\end{tabular}}} \\ \cline{2-5} 
 & \hspace{12pt} \textbf{$\uparrow \uparrow$} \hspace{12pt} & \textbf{$\downarrow \uparrow$} & \hspace{12pt} \textbf{$\uparrow \uparrow$} \hspace{12pt} & \textbf{$\downarrow \uparrow$} \\ \hline
injected by 8 & 1.8 & -1.8 & 1.6 & -1.6 \\ \hline
injected by 9 & 2.1 & 2.1 & 2.5 & 2.5 \\ \hline
\textbf{Total $\mu_{s}$} & \textbf{3.9} & \textbf{0.3} & \textbf{4.1} & \textbf{0.9} \\ \hline
\end{tabular}
\caption{Spin accumulation $\mu_{s}$ in graphene, beneath the contacts, in the two-terminal spin-valve geometry at bias $I=+20~\mu$A. The arrows $\uparrow \uparrow$ ($\downarrow \uparrow$) represent the parallel (anti-parallel) orientation of the magnetization of contacts 8 and 9 respectively, from left to right. }
\label{Table_spin_accumulation}
\end{table}
\section{Methods: Device preparation}
A fully encapsulated hBN/graphene/hBN heterostructure is prepared via a dry pickup transfer method developed in our group \cite{44_Zomer2014_APL}. The graphene flake is exfoliated from a bulk HOPG (highly oriented pyrolytic graphite) ZYA grade crystal (supplier: SPI) onto a pre-cleaned SiO$_{\text{2}}$/Si substrate ($t_{\text{SiO}_{\text{2}}}$=300 nm). A single layer is identified  via the optical contrast analysis. Boron nitride flakes (supplier: HQ Graphene) are exfoliated onto a different SiO$_{\text{2}}$/Si substrate ($t_{\text{SiO}_{\text{2}}}$=90 nm) from small hBN crystals ($\sim$ 1 mm). The  thickness of the desired hBN flake is characterized via the Atomic Force Microscopy (AFM).  For the stack preparation, a  bilayer-hBN (2L-hBN) flake on a SiO$_{\text{2}}$/Si is brought in contact with a viscoelastic PDMS (polydimethylsiloxane) stamp which has a polycarbonate (PC) film attached to it in a transfer stage arrangement. When the sticky PC film comes in a contact with a 2L-hBN flake, the flake is picked up by the PC film. A single layer graphene (Gr) flake, exfoliated onto a different SiO$_{\text{2}}$/Si substrate is aligned with respect to the already picked up 2L-hBN flake in the transfer stage. When the graphene flake is brought in contact with the 2L-hBN flake on the PC film, it is picked up by the 2L-hBN flake due to van der Waals force between the flakes. In the last step, the 2L-hBN/Gr assembly is aligned on top of a 10 nm thick-hBN flake on another SiO$_{\text{2}}$/Si substrate and brought in contact with the flake. The whole assembly is heated at an elevated temperature $\sim$ 150$^{\circ}$C and the PC film with the 2L-hBN/Gr is released onto the thick-hBN flake. The PC film is dissolved by putting the stack in a chloroform solution for three hours at room temperature. Then the stack is annealed at 350$^{\circ}$C for 5 hours in an Ar-H$_{\text{2}}$ environment for removing the polymer residues.

The electrodes are patterned via the electron beam lithography on the PMMA (poly(methyl methacrylate)) coated 2L-hBN/Gr/hBN stack. Following the development procedure, which selectively removes the PMMA exposed to the electron beam, 65 nm thick ferromagnetic (FM) cobalt electrodes are deposited on top of the 2L-hBN tunnel barrier for the spin polarized electrodes via electron-beam evaporation.  Vacuum  pressure is maintained at $1 \times 10^{-7}$ mbar during the deposition. To prevent the oxidation of the cobalt, the ferromagnetic electrodes are covered with a 3 nm thick aluminum layer. The material on top of the unexposed polymer is removed via the lift-off process in hot acetone solution at 50$^{\circ}$C, leaving only the contacts in the desired area.

 
\bibliography{Manuscript_2_2016_Draft5_Main}

\section{Acknowledgements}
We kindly acknowledge J.G. Holstein, H.M. de Roosz, H. Adema and T.J. Schouten for the technical assistance. We thank Prof. T. Banerjee for useful discussions and J. Ingla-Ayn\'{e}s for providing the device with TiO$_2$ barrier. The research leading to these results has received funding from the European-Union Graphene Flagship (grant no. 604391) and supported by the Zernike Institute for Advanced Materials and Nederlandse Organisatie voor Wetenschappelijk (NWO,
Netherlands).

\section{Author contributions}
M.G., S.O., and B.J.v.W. conceived the experiments. M.G. carried out the sample fabrication and measurements. M.G., S.O., and B.J.v.W. carried out the analysis and wrote the manuscript. All authors discussed the results and the manuscript.

\section{Competing financial interests}
\starttocentries 
The authors declare no competing financial interests.

\starttocentries 

\newpage




\beginsupplement

\begin{center}
 \textbf{\large SUPPLEMENTARY INFORMATION} \\ 
\vspace*{1 mm}
 \textbf{\large  Bias induced up to 100\% spin-injection and detection polarizations in ferromagnet/bilayer-hBN/graphene/hBN heterostructures} \\
 \vspace*{1 mm}
 M. Gurram, S. Omar, and B.J. van Wees	\\
\vspace*{1 mm}
Physics of Nanodevices, Zernike Institute for Advanced Materials, University of Groningen, The Netherlands \\
\vspace*{1 mm}
(\today)
\end{center}

\tableofcontents	

\stoptocentries  
\starttocentries  
\newpage

\section{Device characterization}
\begin{figure}[!htbp]
\centering
 \includegraphics[width=\columnwidth,trim= 0in 0in 0in 0in,clip]{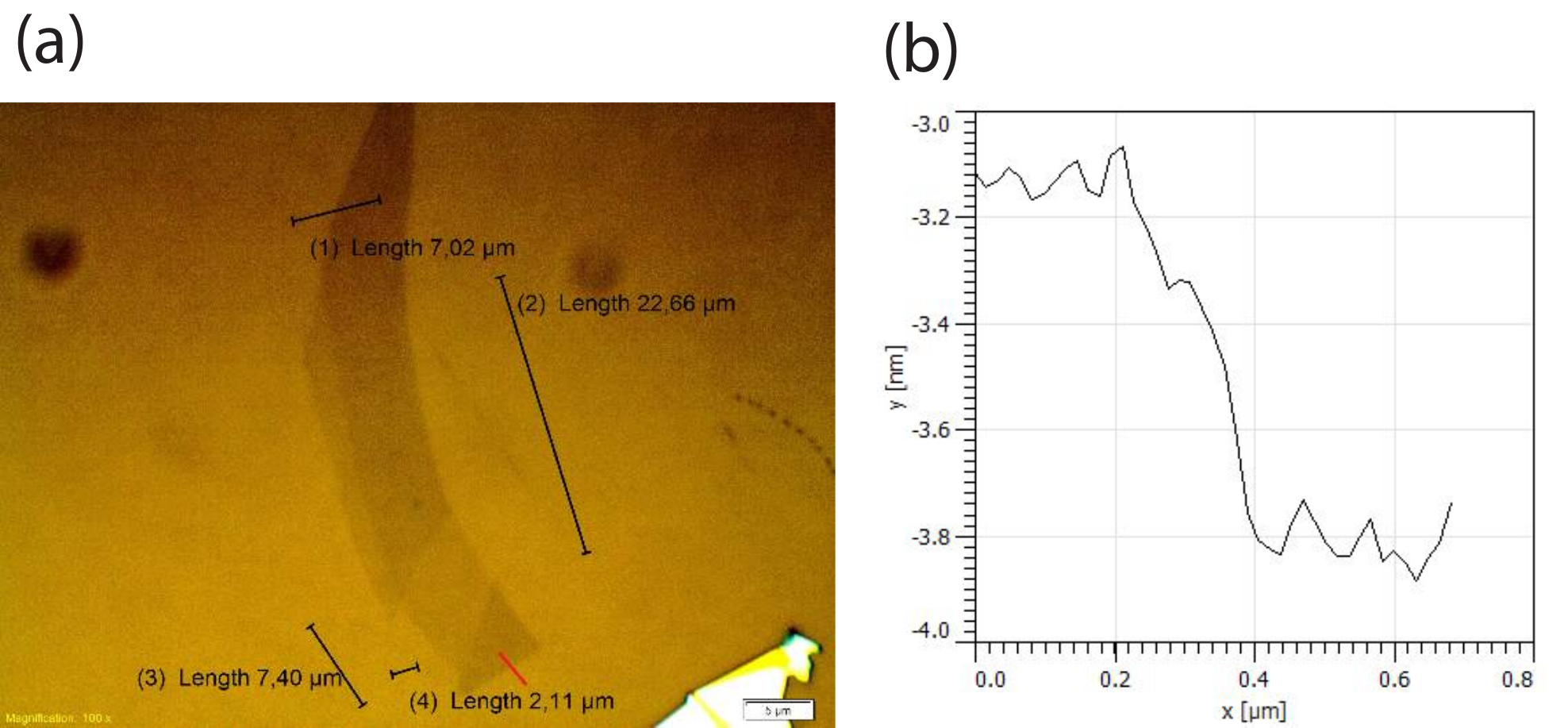}
 \caption{(\textbf{a}) An optical microscopic image of the  hBN tunnel barrier flake on a Si/SiO$_{\text{2}}$ substrate ($t_{SiO_2}$= 90 nm) where the lighter contrast regions indicate the single-layer hBN. (\textbf{b}) An AFM height profile of the 2L-hBN corresponding to the red line drawn in \textbf{a}, showing a thickness value $\sim$ 0.7 nm.}
 \label{AFM}
\end{figure}

An optical microscopy image of the 2L-hBN flake and its AFM thickness measurement is shown in Fig.~\ref{AFM}. Charge and spin transport measurements in graphene are performed using low-frequency (21 Hz) lock-in measurements. All measurements are performed in vacuum  ($\sim 1 \times 10^{-7}$ mbar) at room temperature. In order to eliminate the effect of the contact resistances, the graphene resistivity was characterized using a four-terminal local geometry by applying an AC current between contacts 1-13 and measuring the voltage drop across a pair of contacts in between 1 and 13 (see Fig.~1b of the main text). The square resistance $R_{\text{sq}}$ of graphene is consistently found to be  $\sim 400~\Omega $ for different regions, suggesting that the background doping profile is uniform in the fully encapsulated graphene flake.

\begin{figure}[!htbp]
\centering
 \includegraphics[width=(\columnwidth),trim= 0in 0in 0in 0in,clip]{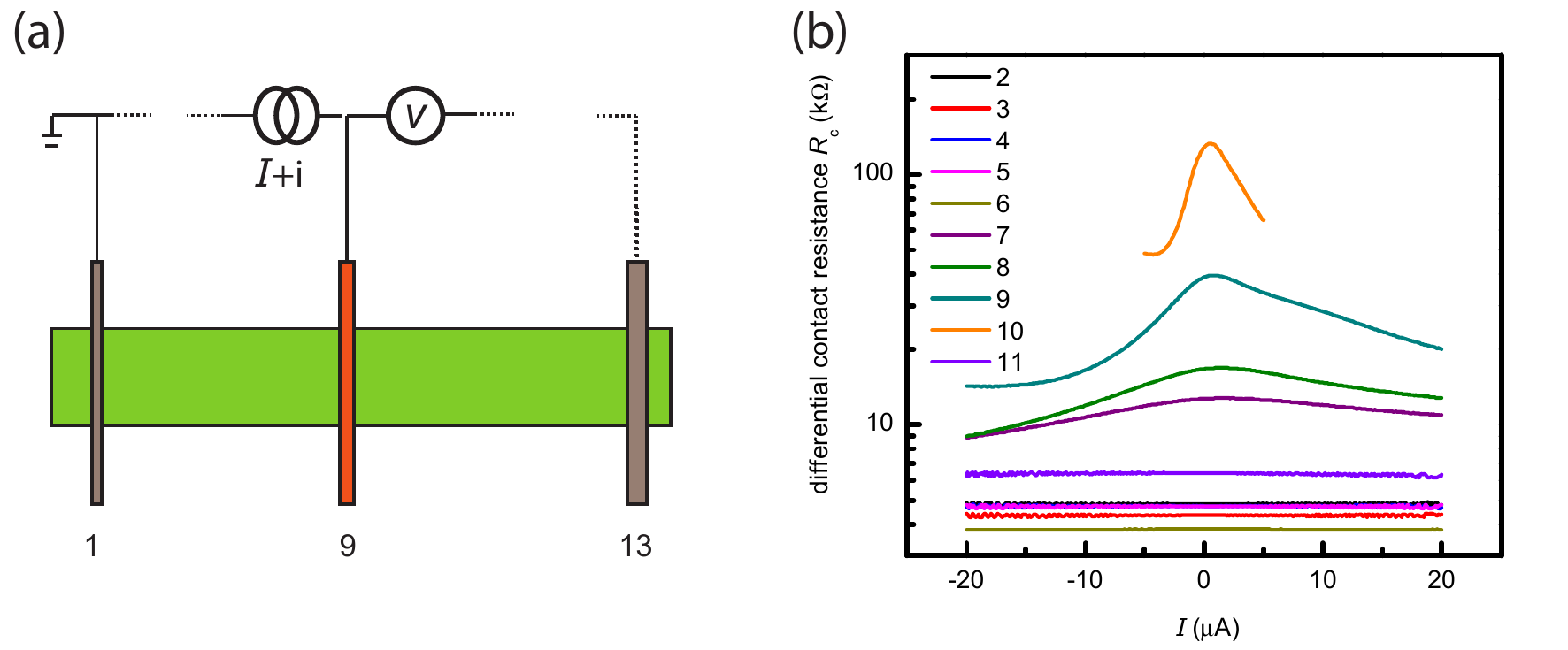}
 \caption{Electrical characterization of the contacts (see section-II).}
 \label{Fig_SI_2_Rc} 
\end{figure}

The differential contact resistances $R_{c}$(=$dV/dI$) of the cobalt/2L-hBN/graphene interface were characterized using a three-terminal connection scheme. For example, to determine the differential resistance of contact 9, a small and fixed AC current ($i$) along with a DC current bias ($I$) is applied between contacts 9-1, and a differential (AC) voltage is measured between 9-13 (Fig.~\ref{Fig_SI_2_Rc}a) while sweeping the DC bias $I$. The resulting data is plotted in the Fig.~\ref{Fig_SI_2_Rc}b. The non-linear behaviour of the high resistive contacts is an indication of the tunneling nature of the 2L-hBN tunnel barrier, whereas the nearly constant differential resistance in the applied bias range, is a characteristic of a transparent (ohmic) contact. For our sample, the differential contact resistances are in the range of 4 - 130 k$\Omega$, and the data is summarized in a table~\ref{Rc_Table}.


\begin{table}[!htbp]
\centering
\label{Rc_Table}	
\begin{tabular}{
>{\columncolor[HTML]{FFCCC9}}c ccccc}
\hline
\cellcolor[HTML]{9B9B9B}\textbf{Contact \#} & \cellcolor[HTML]{9B9B9B}\textbf{\begin{tabular}[c]{@{}c@{}}R$_{c}$ (k$\Omega$)\\(at V=0) \end{tabular}} & \cellcolor[HTML]{9B9B9B}\textbf{\begin{tabular}[c]{@{}c@{}}Width of contact (L) \\ ($\mu$m)\end{tabular}} & \cellcolor[HTML]{9B9B9B}\textbf{\begin{tabular}[c]{@{}c@{}}R$_{c}$*Area \\ (k$\Omega.\mu m^2$)\end{tabular}} & \cellcolor[HTML]{9B9B9B}\textbf{$R_{c}/R_{\lambda}$} & \cellcolor[HTML]{9B9B9B}\textbf{\begin{tabular}[c]{@{}c@{}}No. of hBN layers \\of the barrier\end{tabular}} \\ \hline
\textbf{2} & 4.82 & 0.25 & 3.61 & 6.23 & 1 \\
\textbf{3} & 4.34 & 0.20 & 2.60 & 5.61 & 1 \\
\textbf{4} & 4.74 & 0.17 & 2.41 & 6.12 & 1 \\
\textbf{5} & 4.73 & 0.20 & 2.83 & 6.11 & 1 \\
\textbf{6} & 3.82 & 0.40 & 4.58 & 4.93 & 2 \\
\textbf{7} & 12.7 & 0.35 & 13.3 & 16.4 & 2 \\
\textbf{8} & 16.7 & 0.25 & 12.5 & 21.6 & 2 \\
\textbf{9} & 38.8 & 0.15 & 17.5 & 50.3 & 2 \\
\textbf{10} & 128 & 0.20 & 77.1 & 166 & 2 \\
\textbf{11} & 6.41 & 0.40 & 7.69 & 8.28 & 2 \\ 
\textbf{12} & 10.2 & 0.35 & 12.2 & 13.2 & 2 \\\hline
\end{tabular}
\caption{A summary of all the used contacts. Here $R_{\lambda}=R_{\text{sq}}\lambda_{\text{s}}/W = 773~\Omega$ is the spin resistance of the graphene flake with the width $W$ = 3 $\mu$m, spin-relaxation length $\lambda_{\text{s}}$ = 5.8 $\mu$m, and R$_{\text{sq}}$ $\sim$ 400 $\Omega$. The number of hBN layers is determined from the optical contrast analysis of the optical microscopic images and the AFM measurements.}
\end{table}

\section{Expressions for spin-injection and detection polarizations, and two-terminal local spin-signal}
\stoptocentries
\subsection{Injection polarization}
We derive an analytical expression for a DC/AC spin injection and detection polarizations. In our measurements, we observe that the measured polarization depends on the applied DC current bias ($I$) across the contact. For the DC current injection, the DC polarization of an injector contact $P_{in}$ is defined as:

\begin{equation}
P_{in}(I)=\frac{I_{\text{s}}}{I}
 \label{dc polarization}
\end{equation}
where $I_{\text{s}}$ is the DC spin current and $I$ is the injected DC charge current.  
Similarly, the AC (differential) polarization of the injection contact $p_{in}$, in the presence of a DC bias current $I$, is defined as: 

\begin{equation}
p_{in}(I)=\frac{i_{\text{s}}}{i}
 \label{ac polarization}
\end{equation}
 where $i_{\text{s}}$ is the AC spin current and $i$ is the injected AC charge current.
 
 In our experiment, we apply a DC current at the injector contact along with a small and fixed magnitude of the AC current. The total injected spin current can be represented as:
 
 \begin{equation}
 I_{\text{s}}(I+i)=P_{in}(I+i)\times (I+i)
  \label{total spin current}
 \end{equation}

 Eq.~\ref{total spin current} can be expanded in to a Taylor series. For a small and fixed AC current $i$, the second order terms can be neglected and the expression can be rewritten as:
 
 \begin{equation}
 I_{\text{s}}(I)+ \left.{\left(\frac{dI_{\text{s}}}{dI}\right)}\right|_{I} \times i = P_{in}(I)\times I+ \left\lbrace P_{in}(I) + \left.{\left(\frac{dP_{in}}{dI}\right)}\right|_{I}\times I\right\rbrace  \times i
  \label{total spin current taylor}
 \end{equation}

 The AC (differential) polarization can then be written as: 
  
  \begin{equation}
   p_{in}(I)= \frac{dI_{\text{s}}}{dI}=\frac{i_{\text{s}}}{i}= P_{in}(I)+ \left.{\left(\frac{dP_{in}}{dI}\right)}\right|_{I} \times I
   \label{ac injection polarization}
  \end{equation}
  
   Eq.~\ref{ac injection polarization} can be used for a consistency check between the measured $p_{in}$ and $P_{in}(I)$ (Fig.~\ref{Pac_meas_analytical}).  

In our case, we observe that $p_{in}$ approximately scales linearly with bias $I$, implying that $\frac{dP_{in}}{dI}\sim$ constant. Eq.~\ref{ac injection polarization} then gives $P_{in}$ $\approx$ $\frac{1}{2}p_{in}(I)$.

\subsection{Detection polarization}  
  The spin-detection polarization is defined as a voltage measured at the detector due to the spin accumulation underneath the detector contact. 
  A charge current $\triangle I$ will flow in the ferromagnet via a spin-charge coupling due to a change in the spin accumulation $\triangle \mu_{\text{s}}$ underneath the detector:
  \begin{equation}
  \triangle I= \triangle\mu_{s}(\frac{dI_{\uparrow}}{dV}-\frac{dI_{\downarrow}}{dV})
   \label{detection 1}
  \end{equation}

  where the net spin accumulation $\mu_{s}$ is the splitting of spin chemical potentials spin-up $\mu_{\uparrow}$ and spin-down $\mu_{\downarrow}$, i.e., ($\mu_{\uparrow}-\mu_{\downarrow}$)/2.
  Note that Eq.~\ref{detection 1} holds under the condition of independent spin channels.
  For a fixed current bias $I$ at the detector, to compensate for $\triangle I$, the change in the voltage $\triangle V$ at the detector will give rise to a change in the charge current $\triangle I$ in the opposite direction:
  \begin{equation}
   \triangle I=\triangle V (\frac{dI_{\uparrow}}{dV}+\frac{dI_{\downarrow}}{dV})
   \label{detection 2}
\end{equation}

  Solving Eq.~\ref{detection 1} and Eq.~\ref{detection 2} leads to:
\begin{equation}
 \frac{\triangle V}{\triangle \mu_{\text{s}}}=\frac{\frac{dI_{\uparrow}}{dV}-\frac{dI_{\downarrow}}{dV}}{\frac{dI_{\uparrow}}{dV}+\frac{dI_{\downarrow}}{dV}}=\frac{dI_{\uparrow}-dI_{\downarrow}}{dI_{\uparrow}+dI_{\downarrow}}=\frac{dI_{\text{s}}}{dI}
 \label{detection_polarization}
\end{equation}
Since the spin accumulations underneath the detector contacts are generally small, this equation is valid for both the DC detector polarization $P_d$ and the differential detector polarization $p_d$, i.e.,
\begin{equation}
 P_d(I)=p_d(I)=\frac{dI_{\text{s}}}{dI}=p_{in}(I)
 \label{detection_polarization_2}
\end{equation}

Note that electrons can only inject one spin $\hbar$/2 (up or down), which implies that $P_{in}(I)$ is restricted below $\pm$ 100\%. However, this does not hold for the differential injection polarization $p_{in}$ as well as detection polarizations $p_{d}(I)$ and $P_{d}(I)$ which can in principle exceed $\pm$100\% in case of applied bias. Note however that when a detector is biased, it will also inject spins resulting in a spin accumulation underneath the detector. When the detector is fully spin polarized, the spin induced voltage $V$ cannot exceed the total spin accumulation $\pm \mu_{s, total}/e$ (due to injector and detector). As it can be seen from the Table I of the main text, this condition is always satisfied, since the sum of the spin induced voltages cannot be larger than $\mu_{s, total}/e$ = (3.9+4.1)/e = 8 mV which is in agreement with the signal in Figure~5d of the main text.
\subsection{Two-terminal local spin signals}
\starttocentries
We can calculate the bias-dependent two-terminal spin signal, provided the spin injection and detection polarizations are known. For the two-terminal measurements, the injector and detector are both biased with the same DC current $I$ but they are biased with opposite polarity. The two-terminal DC spin signal $\triangle V_{\text{2t}}^{\text{DC}}$ between contacts 8 and 9 (See Fig. 5 in the main text) can be written as:
\begin{equationsenviron}
 \triangle V_{\text{2t}}^{\text{DC}} = I\times[P_{in}^{9}(I)P_{d}^{8}(-I) + P_{in}^{8}(-I)P_{d}^{9}(I)]\times \frac{R_{\text{sq}}\lambda_{\text{s}}}{W} \times e^{-\frac{L}{\lambda_{\text{s}}}}
 \label{dc spin signal}
\end{equationsenviron} 
which is equal to $V_{2t}^{\uparrow \uparrow}(I) - V_{2t}^{\downarrow \uparrow}(I)$, the difference in two-terminal DC voltage signal $V_{\text{2t}}^{\text{DC}}$ when the magnetization configuration of contacts 8 and 9 changes between parallel($\uparrow\uparrow$) and anti-parallel($\downarrow\uparrow$) (see the main text).

Similarly, the two-terminal differential spin signal $\triangle R_{\text{2t}}^{\text{AC}}$ between contacts 8 and 9 (See Fig. 5 in the main text) can be written as:
\begin{equationsenviron}
 \triangle R_{\text{2t}}^{\text{AC}} = [p_{in}^9(I)p_d^8(-I) + p_{in}^8(-I)p_d^9(I)]\times \frac{R_{\text{sq}}\lambda_{\text{s}}}{W} \times e^{-\frac{L}{\lambda_{\text{s}}}}
 \label{ac spin signal SI}
\end{equationsenviron} 

which is equal to $R_{2t}^{\uparrow \uparrow}(I) - R_{2t}^{\downarrow \uparrow}(I)$, the difference in the two-terminal differential signal $R_{\text{2t}}^{\text{AC}}$ when the magnetization configuration of contacts 8 and 9 changes between parallel($\uparrow\uparrow$) and anti-parallel($\downarrow \uparrow$).

Here $L$ is the separation between the contacts 8 and 9. $p_{in}$ and $p_d$ are obtained by following the procedure explained in the next section. 
 
 
\section{Determining the bias dependent spin-injection polarizations from non-local spin signals} \label{sec_P_inj_ac} \label{section_pin_analysis}
In a typical non-local spin-valve measurement, a differential voltage signal $v_{nl}$, measured by a detector contact 'd' with differential detection polarization $p_{\text{d}}$, located at a distance L from an injector contact 'in' with differential injection polarization $p_{\text{in}}$, is given by
\begin{equation}
  v_{nl}= \frac{iR_{\text{sq}}\lambda_{\text{s}}}{2W} p_{\text{in}}p_{\text{d}}e^{-L/\lambda_{\text{s}}}
 \label{Vid}
\end{equation}
where $R_{\text{sq}}$ is the square resistance of graphene, $\lambda_{\text{s}}$ is the spin relaxation length in graphene and $W$ is the width of the graphene flake.

\begin{figure}[!htbp]
\centering
 \includegraphics[width=\columnwidth,trim= 0in 0in 0in 0in,clip]{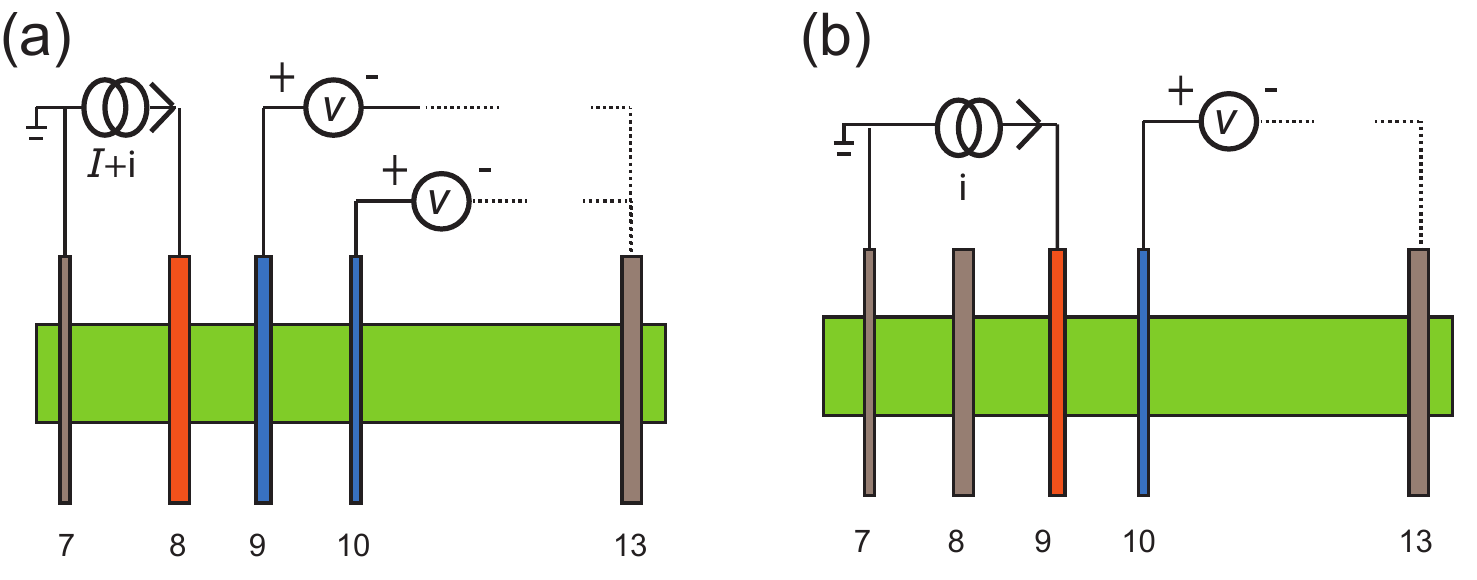}
 \caption{Schematics of the measurement configurations for determining the spin-injection polarization of the contact 8. (\textbf{a}) Measuring non-local spin signals as a function of bias on injector contact. (\textbf{b}) Measuring unbiased non-local spin signal with the two detector contacts 9 and 10.}
 \label{non_local}
\end{figure}

Consider a group of five contacts 7, 8, 9, 10, and 13 in Fig.~\ref{non_local}a, where the current ($I + i$) is injected through a ferromagnet in contact 8 and extracted through 7, and the total differential spin accumulation is detected as a non-local differential voltage, using a low-frequency lock-in detection scheme, between the contacts 9 and 13 $v_{\text{nl}}^{9-13}$.

The non-local voltage measured with the magnetization of the contacts 7, 8, 9, and 13 are aligned in one direction (say $\uparrow \uparrow \uparrow \uparrow $) is given by,

\begin{equation}
 v_{\text{nl}}^{9-13}(\uparrow \uparrow \uparrow \uparrow)= \frac{iR_{sq}\lambda_{s}}{2W} \left[ p_{9}\left( p_{8}e^{-L_{8-9}/\lambda_{\text{s}}} - p_{7}e^{-L_{7-9}/\lambda_{\text{s}}} \right)  -  p_{13}\left( p_{8}e^{-L_{8-13}/\lambda_{\text{s}}} - p_{7}e^{-L_{7-13}/\lambda_{\text{s}}} \right)  \right] 
 \label{Vac}
\end{equation}

In our measurements, the outer detector 13 is far enough from the injectors ($L_{7-13}$, $L_{8-13}$ $>$ 2-3*$\lambda_{s}$) to not to detect any spin signal and serves as a reference detector for the rest of the analysis. So, the non-local differential resistance $R_{nl}$ = $v_{\text{nl}}/i$ detected by 9 due to injection from 7 and 8 is given by

\begin{equation}
 R_{\text{nl}}^{\uparrow \uparrow \uparrow}= \frac{R_{sq}\lambda_{s}}{2W} \left[ p_{9}\left( p_{8}e^{-L_{8-9}/\lambda_{\text{s}}} - p_{7}e^{-L_{7-9}/\lambda_{\text{s}}} \right)   \right] 
 \label{V123_P}
\end{equation}

In a spin-valve measurement, when the magnetization of one of the contact (say, 8) switches, the resulting non-local resistance can be written as,
\begin{equation}
 R_{\text{nl}}^{\uparrow \downarrow \uparrow}= \frac{R_{sq}\lambda_{s}}{2W} \left[ p_{9}\left( -p_{8}e^{-L_{8-9}/\lambda_{\text{s}}} - p_{7}e^{-L_{7-9}/\lambda_{\text{s}}} \right)   \right] 
 \label{V123_AP}
\end{equation}

The detected signals in the equations Eq.~\ref{V123_P} and Eq.~\ref{V123_AP} include the contribution of spin signal from the outer injector 7 (second term of the expressions) as well as some field independent background signal.

Since the only change in equations Eq.~\ref{V123_P} and Eq.~\ref{V123_AP} is due to contact 8, the non-local spin signal measured by 9 corresponding to the spin accumulation created only by 8 is obtained from
\begin{equation}
\Delta R_{\text{nl}}^{8-9} = \frac{ R_{\text{nl}}^{\uparrow \uparrow \uparrow} - R_{\text{nl}}^{\uparrow \downarrow \uparrow} }{2} = \frac{R_{\text{sq}}\lambda_{s}}{2W} \left[ p_{9}\left( p_{8}e^{\frac{-L_{8-9}}{\lambda_{\text{s}}}} \right) \right]  
 \label{dRnl_89}
\end{equation}

As explained above, one can determine the spin signal measured via inner detector contact 9 correspond to the spin injection through inner injector contact 8 as given by Eq.~\ref{dRnl_89}.
Further, as shown in Fig.~\ref{non_local}(b), we can simultaneously measure the spin signal via inner detector contact 10 corresponding to the spin injection through inner injector contact 8, given by
\begin{equation}
 \Delta R_{\text{nl}}^{8-10} = \frac{R_{\text{nl}}^{\uparrow \uparrow \uparrow} - R_{\text{nl}}^{\uparrow \downarrow \uparrow}}{2} = \frac{R_{sq}\lambda_{s}}{2W} \left[ p_{10} \left( p_{8}e^{\frac{-L_{8-10}}{\lambda_{\text{s}}}} \right)  \right]
 \label{dRnl_810}
\end{equation}

The contact polarization of the contacts 9 and 10 can be expressed as a ratio of $\triangle R_{\text{nl}}^{\text{8-9}}$ and $\triangle R_{\text{nl}}^{\text{8-10}}$ i.e.
\begin{equation}
  \frac{p_{\text{9}}}{p_{\text{10}}}=\frac{\triangle R_{\text{nl}}^{\text{8-9}}}{\triangle R_{\text{nl}}^{\text{8-10}}}e^{\frac{-L_{9-10}}{\lambda_{\text{s}}}}
  \label{polarization ratio}
\end{equation}

In order to determine the unbiased values of detector polarizations $p_{\text{9}}$ and $p_{\text{10}}$, we need one more equation with these variables which is obtained by measuring $\triangle R_{\text{nl}}$ between 9 and 10, by applying only an AC injection current between contacts 7 and 9 and measuring a non-local voltage between 10 and 13. The effect of the outer injector contact 7 is subtracted using the procedure described above(see equations~\ref{Vac} - \ref{dRnl_89}). Now we obtain:
\begin{equation}
 \Delta R_{\text{nl}}^{9-10} = \frac{R_{sq}\lambda_{s}}{2W} \left[ p_{10} \left( p_{9}e^{\frac{-L_{9-10}}{\lambda_{\text{s}}}} \right)  \right]
 \label{dRnl_910}
\end{equation}

We can obtain the product $p_{\text{9}}\times p_{\text{10}}$ from Eq.~\ref{dRnl_910} and the ratio $\frac{p_{\text{9}}}{p_{\text{10}}}$ from Eq.~\ref{polarization ratio} and thus determine the unbiased polarizations $p_{\text{9}}$ and $p_{\text{10}}$. 

\begin{figure}[!htbp]
\centering
 \includegraphics[width=\columnwidth,trim= 0in 0in 0in 0in,clip]{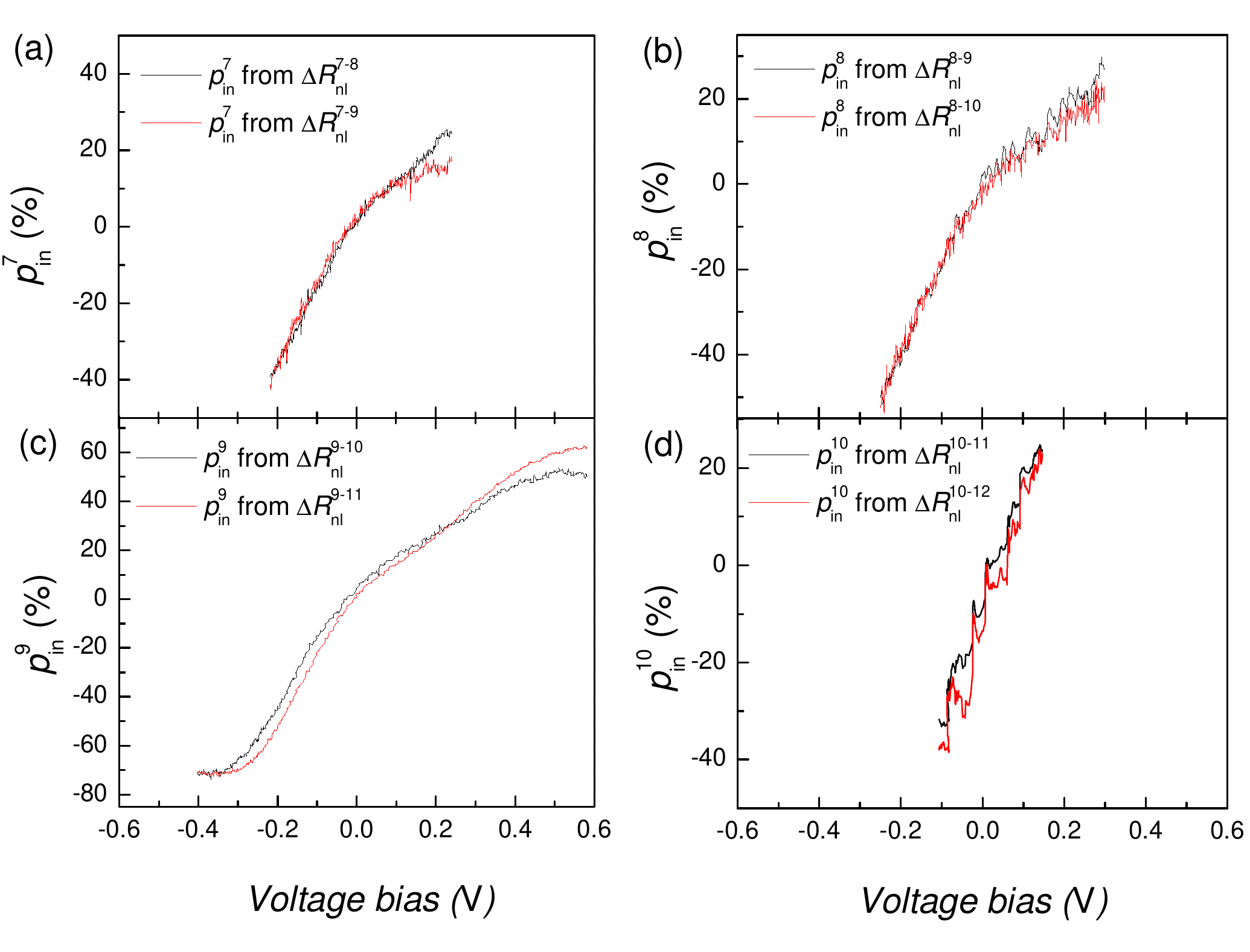}
 \caption{Differential spin-injection polarization for different injector contacts as a function of DC voltage bias across the injector.  The spin injection into graphene from the FM cobalt is facilitated via a 2L-hBN tunnel barrier, clearly demonstrating the change in the magnitude and the sign of the injector polarization as a function of the bias current $I$ . The injection polarizations of contact 7 $p_{\text{7}}$ in (\textbf{a}), contact 8 $p_{\text{8}}$in (\textbf{b}), contact 9 $p_{\text{9}}$ in (\textbf{c}), and contact 10 $p_{\text{10}}$ in (\textbf{d}) are shown.}
 \label{injector_polarization}
\end{figure}

Using the unbiased polarization values of detectors obtained from Eq.~\ref{polarization ratio} and Eq.~\ref{dRnl_910}, we can determine the bias dependent polarization of the injector contact 8 from the two non-local spin signals measured via contacts 9 (Eq.~\ref{dRnl_89}) and contact 10 (Eq.~\ref{dRnl_810}), independently. The resulting differential spin-injection polarization of contact 8 is plotted in figure~\ref{injector_polarization}(b).

The above procedure is repeated with three more different groups of contacts to determine the differential polarization of injection contacts, and the results are plotted in the figure~\ref{injector_polarization}(a-d). The results are also summarized in the table~\ref{table_p7p8p9}.

\begin{table}[!htbp]
\centering
\resizebox{\textwidth}{!}{%
\begin{tabular}{@{}cccccccccc@{}}
\toprule
 &  & \multicolumn{2}{c}{\cellcolor[HTML]{C0C0C0}{\color[HTML]{000000} At V = 0}} &  & \multicolumn{2}{c}{\cellcolor[HTML]{C0C0C0}At V = +V$_{max}$} &  & \multicolumn{2}{c}{\cellcolor[HTML]{C0C0C0}At V = -V$_{max}$} \\ \midrule
\rowcolor[HTML]{C0C0C0} 
\cellcolor[HTML]{FFCCC9}Set of contacts & \begin{tabular}[c]{@{}c@{}}Injector-detector \\ ($in-d$)\end{tabular} & {\color[HTML]{FE0000} \begin{tabular}[c]{@{}c@{}}$p_{in}$\\ (\%)\end{tabular}} & {\color[HTML]{3531FF} \begin{tabular}[c]{@{}c@{}}$p_{d}$\\ (\%)\end{tabular}} &  & \begin{tabular}[c]{@{}c@{}}$\Delta R_{nl}^{in-d}$\\ ($\Omega$)\end{tabular} & {\color[HTML]{FE0000} \begin{tabular}[c]{@{}c@{}}$p_{in}$\\ (\%)\end{tabular}} &  & \begin{tabular}[c]{@{}c@{}}$\Delta R_{nl}^{in-d}$\\ ($\Omega$)\end{tabular} & {\color[HTML]{FE0000} \begin{tabular}[c]{@{}c@{}}$p_{in}$\\ (\%)\end{tabular}} \\
\cellcolor[HTML]{FFCCC9} & 7-8 &  & $p_{8}$ = -2.0 &  & -1.5 & 24.5 &  & 2.3 & -38.5 \\
\multirow{-2}{*}{\cellcolor[HTML]{FFCCC9}7-8-9} & 7-9 & \multirow{-2}{*}{$p_{7}$ = 1.4} & $p_{9}$ = 1.1 &  & 0.5 & 17.3 &  & -1.1 & -42.6 \\
\cellcolor[HTML]{FFCCC9} &  &  &  &  &  &  &  &  &  \\
\cellcolor[HTML]{FFCCC9} & 8-9 &  & $p_{9}$ = 1.3 &  & 1.2 & 26.9 &  & -1.9 & -50.0 \\
\multirow{-2}{*}{\cellcolor[HTML]{FFCCC9}8-9-10} & 8-10 & \multirow{-2}{*}{$p_{8}$ = -2.3} & $p_{10}$ = 3.0 &  & 1.6 & 22.9 &  & -3.8 & -52.6 \\
\cellcolor[HTML]{FFCCC9} &  &  &  &  &  &  &  &  &  \\
\cellcolor[HTML]{FFCCC9} & 9-10 &  & $p_{10}$ = 2.4 &  & 3.7 & 51.3 &  & -5.2 & -71.0 \\
\multirow{-2}{*}{\cellcolor[HTML]{FFCCC9}9-10-11} & 9-11 & \multirow{-2}{*}{$p_{9}$ = 4.3} & $p_{11}$ = 3.2 &  & 3.9 & 61.8 &  & -4.5 & -70.8 \\
\cellcolor[HTML]{FFCCC9} &  &  &  &  &  &  &  &  &  \\
\cellcolor[HTML]{FFCCC9} & 10-11 &  & $p_{11}$ = 3.2 &  & 1.9 & 23.2 &  & -2.6 & -31.6 \\
\multirow{-2}{*}{\cellcolor[HTML]{FFCCC9}10-11-12} & 10-12 & \multirow{-2}{*}{$p_{10}$ = -1.7} & $p_{12}$ = 2.0 &  & 0.9 & 23.1 &  & -1.5 & -37.9 \\ \cmidrule(lr){2-2} \cmidrule(l){4-10} 
\end{tabular}%
}
\caption{A summary of spin-valve signals and obtained differential spin-injection/detection polarizations. $\Delta R_{nl}^{in-d}(V)$ is the non-local signal from spin-valve data when the injection bias V applied across the injector(in) and measured via detector(d), $p_{in}(V)$ is the differential injection polarization of injector contact at bias V, calculated from the analysis explained in section-IV, and $V_{max(min)}$ is the maximum(minimum) bias applied across the injector. Here, the detector polarization $p_{d}$ at zero bias obtained from following the analysis described in the section-\ref{section_pin_analysis}.}
\label{table_p7p8p9} 
\end{table}

\section{Determining the bias-dependent detector polarizations}
\begin{figure}[!htbp]
\centering
 \includegraphics[width=\columnwidth,trim= 0in 0in 0in 0in,clip]{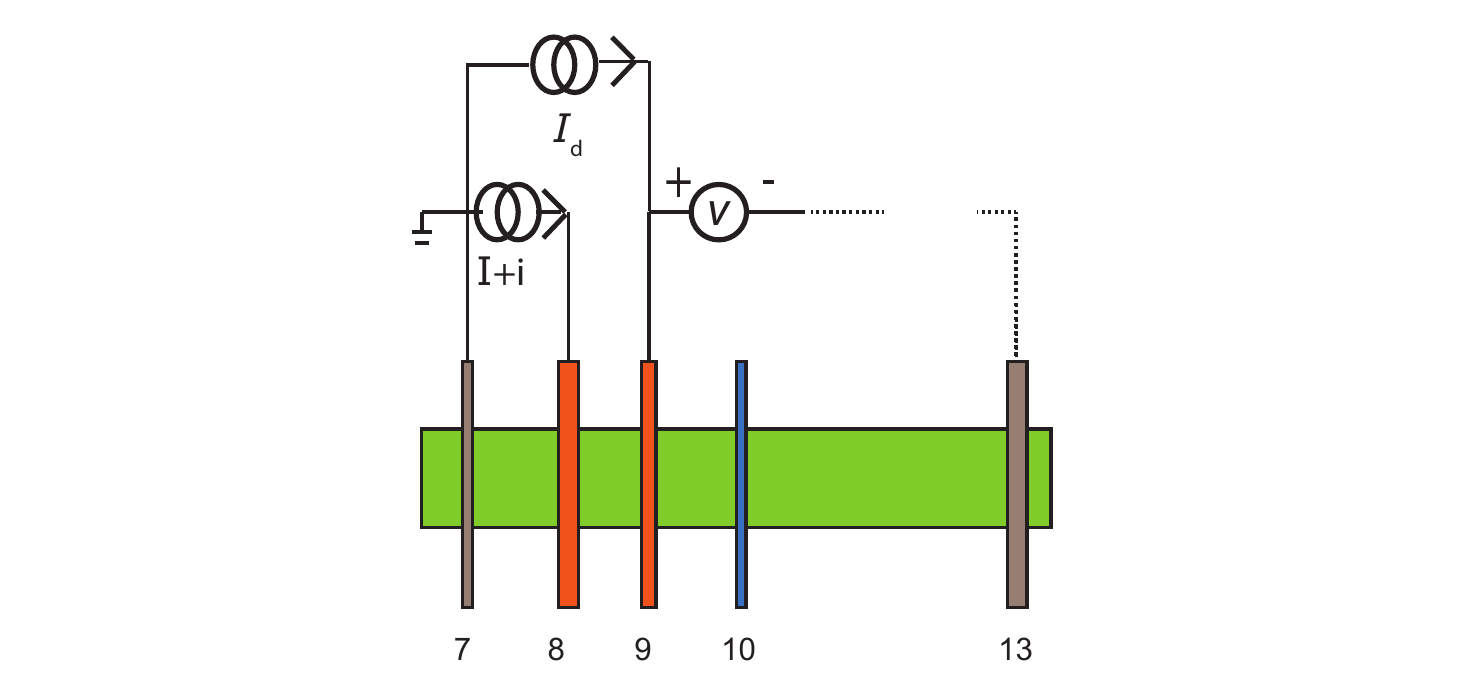}
 \caption{Measurement geometry for biasing the detector to measure the spin-detection polarization}
 \label{detector_polarization}
\end{figure}

In order to measure the bias dependent detector polarization of contact 9, we keep the injector contact 8 at a fixed DC current bias $I$, where $p_{8}(I)$ is known from the previous measurements, and sweep a bias current $I_{d}$ across the detector 9. We apply a fixed $I$ and a small $i$ through the injector electrode 8 and measure a non-local signal at detector 9 via low-frequency lock-in detection method, while sweeping the DC current bias $I_{d}$ across the detector 9 (Fig.~\ref{detector_polarization}). Note that the spin transport is non-local only for the AC measurements. For the DC measurements we have a non-zero charge current and an electric field in the spin transport channel between contacts 8 and 9. A differential non-local signal $\triangle R_{\text{nl}}^{\text{8-9}}$ is measured as a function of detector bias current $I_{d}$ and can be expressed via Eq.~\ref{dRnl_89}. Here, we know the spin-injection polarization $p_{8}(I)$ obtained from the previous measurements (section-IV) and can extract the spin detection polarization as a function of the bias current $I_{d}$ using Eq.~\ref{dRnl_89} (see the main text). 

\section{Differential polarization from DC polarization}
\begin{figure}[!htbp]
\centering
 \includegraphics[width=(\columnwidth)/2,trim= 0in 0in 0in 0in,clip]{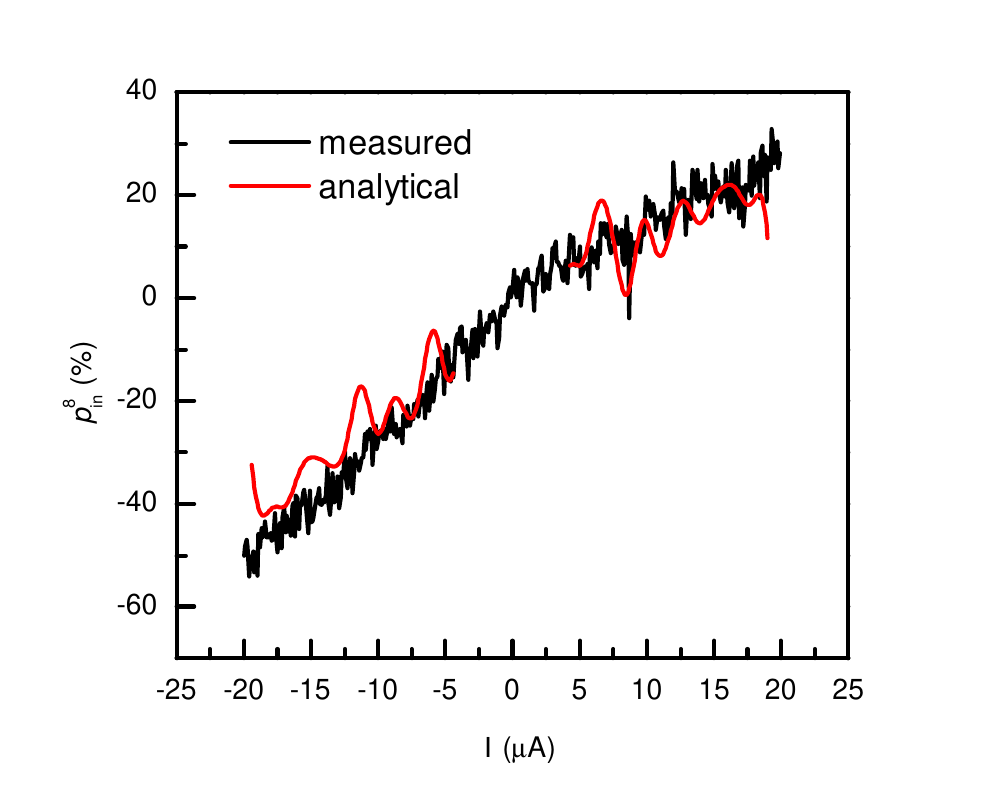}
 \caption{Differential spin-injection polarization of contact 8 obtained from the measurements (black curve) and from the analytical expression Eq.~\ref{ac injection polarization} (red curve).}
 \label{Pac_meas_analytical}
\end{figure}

The differential spin-injection polarization $p_{in}(I)$ can be expressed as the sum of DC injection polarization $P_{in}(I)$, and $\left.{\left(\frac{dP_{in}(I)}{dI}\right)}\right|_{I} I$ (Eq.~\ref{ac injection polarization}). We determine the differential spin-injection polarization of contact 8 $p_{in}^{8}(I)$ as explained in the section IV. A similar analysis is used to determine the DC spin-injection polarization of contact 8 $P_{in}^{8}(I)$ from the DC spin transport measurements where a non-local spin signal is measured via a DC voltmeter. Figure~\ref{Pac_meas_analytical} shows $p_{in}^{8}(I)$ determined both from the measurements and from the analytical expression Eq.~\ref{ac injection polarization}. The measured and the calculated differential polarization ($p_{in}(I)$) are in a good agreement, supporting the consistency of our approach. 

\section{Low interface resistance contacts}


As indicated in the optical microscope picture in Fig.~1b of the main text and Fig.~\ref{AFM}a, a part of the hBN tunnel barrier flake consists of a monolayer(1L)-hBN region. The contacts from 2 to 5, either fully or partially deposited on top of the monolayer region of the tunnel barrier flake, show low interface resistance of $\approx$~4-5 k$\Omega$, whose differential interface resistance $R_{c}$ ($=dV/dI$) is constant as a function of bias (Fig.~\ref{Fig_SI_2_Rc}b).



\begin{figure}[!htbp]
\centering
 \includegraphics[width=(\columnwidth)/2,trim= 0in 0in 0in 0in,clip]{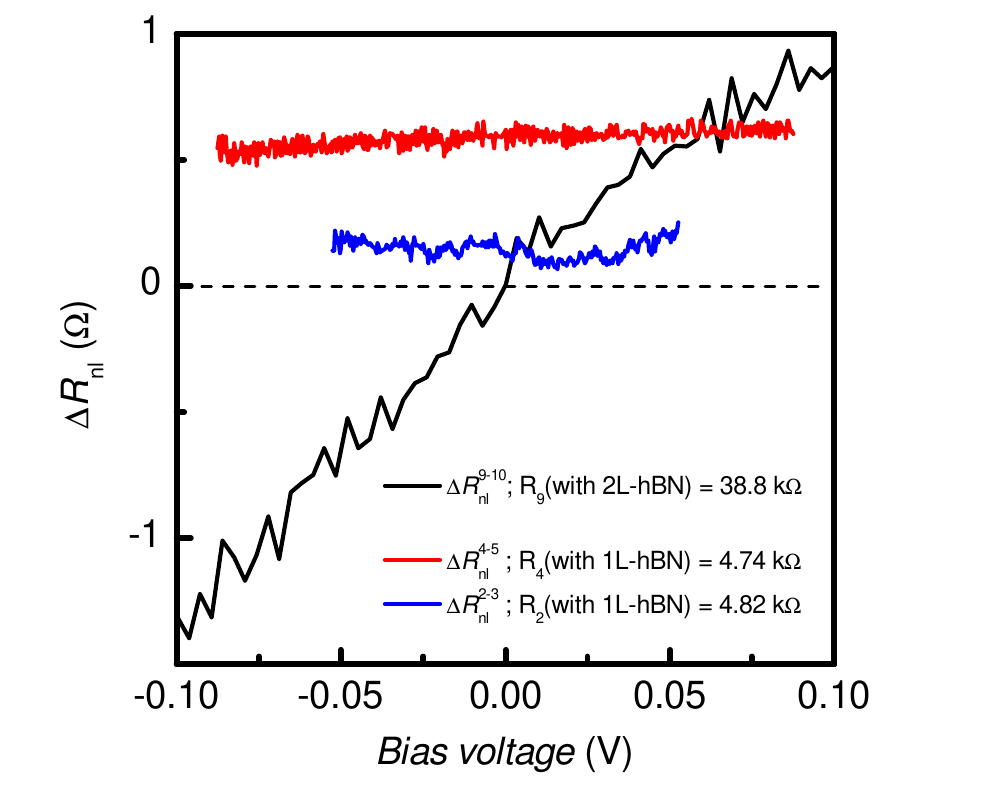}
 \caption{Comparison of spin signals from low and high resistive contacts with 1L-hBN and 2L-hBN barriers.}
 \label{Low_Rc_1L_2LhBN}
\end{figure}

Figure~\ref{Low_Rc_1L_2LhBN} shows the non-local spin-signal corresponding to the spin injection through the low $R_{c}$ contacts 2 and 4, as a function of the applied bias. For a comparison, the spin signal for the high $R_{c}$ contact 9, $\Delta R_{nl}^{\text{9-10}}$ is also shown. For the same range of the applied voltage bias, low $R_{c}$ contacts with 1L-hBN tunnel barriers do not show significant change in the spin signal as well as no sign reversal around zero bias. Whereas the high resistive contacts, for example 9, with 2L-hBN tunnel barriers, show a large modulation as well as change in sign of the non-local spin-signal.

For the used contacts $R_{c}/R_{\lambda} > 5$ (see table~\ref{Rc_Table}) and $L/\lambda \approx 1$, the maximum reduction in $\tau_{s}$ due to the contact induced spin relaxation is within 10\% \cite{50_Thomas2012_PRB}. Here, $R_{\lambda}$=$R_{sq}\lambda_{s}/W$ with square resistance $R_{sq} \sim$ 400 $\Omega$, width W = 3 $\mu$m of graphene, and spin-relaxation length $\lambda_{s}=5.8~\mu$m.

\section{Spin-injection due to heating}
 We use a large value of DC current up to $\pm 20~\mu$A, in order to modulate the spin-injection and -detection polarizations of contacts, which might raise the electron temperature underneath significantly and could inject spins into graphene via a spin-dependent Seebeck effect \cite{vera-marun_spin_2014}. We can roughly estimate the electron temperature in graphene due to Joule heating ($VI \sim$ 10 $\mu$W) at the interface, provided the hBN-SiO$_{2}$ thermal resistance ($R_{\text{th}}$) is known. 
 
 Since the thermal conductivity of hBN ($\kappa \sim$ 380~Wm$^{-1}$K) is 200 times higher than SiO$_{\text{2}}$ ($\kappa \sim$~1.2~Wm$^{-1}$K), the heat flow will be limited by the SiO$_{\text{2}}$ thermal conductivity. The effective contact area is about 1 $\mu m^2$ and in this area, the heat will flow  and spread approximately 1~$\mu$m in the SiO$_{\text{2}}$/Si reservoir. The effective thermal resistance R$_{th}$ of the reservoir will be approximately 3~$\times 10^5$~KW$^{-1}$. 
%
%
%
 An increase in the temperature $\triangle T$ due to heating can be related as:
 \begin{equation}
  \triangle T= {Q} R_{\text{th}}
 \end{equation}
 where ${Q}$ is the heat transport rate i.e., heating at the interface. We obtain $\triangle T \sim$ 3 K on SiO$_{\text{2}}$/Si substrate.  
 
The high value of the DC current will heat up the tunnel junction and could mimic a spin accumulation due to temperature gradient and the spin dependent Seebeck coefficient of the interface \cite{vera-marun_spin_2014}. In our experiments, however, we also demonstrate the modification of the spin-detection polarization along with the spin-injection polarization, which cannot be explained via these effects. Therefore, the effect of heating on the spin transport can be disregarded in our case. 

\section{Carrier density estimation underneath the contact}

In graphene, the carrier density can be estimated from the Einstein relation:

\begin{equation}
 \sigma=\frac{1}{R_{\text{sq}}}=e^2D_{\text{c}}\nu(E_{\text{F}})
 \label{Einstein}
\end{equation}
where $D_{\text{c}}$ is the charge diffusion coefficient, $\nu(E_{\text{F}})$ is the density of states at the Fermi energy $E_{\text{F}}$, which given by the following equation:

\begin{equation}
 \nu(E)=\frac{g_{\text{s}}g_{\text{v}}2\pi |E|}{h^2 v_{\text{F}}^2}
 \label{dos}
\end{equation}
where $g_{\text{s}}=2$ and $g_{\text{v}}$=2, are the spin degeneracy and the valley degeneracy of the electron, respectively, and $v_{\text{F}}= 10^{6}~m/s$, is the Fermi velocity of the electron.
The density of the carriers $n$ can be estimated by integrating Eq.~\ref{dos} from zero to $E_{\text{F}}$:
\begin{equation}
 n=\frac{g_{\text{s}}g_{\text{v}}\pi E_{\text{F}}^2}{h^2 v_{\text{F}}^2}
 \label{carrier}
\end{equation}

Using Eq.~\ref{dos}, Eq.~\ref{carrier}, and Eq.~\ref{Einstein}, $n$ can be obtained from \cite{jozsa_linear_2009}:

\begin{equation}
 n={\left(\frac{hv_{\text{F}}}{R_{\text{sq}}2e^2\sqrt{g_{\text{s}}g_{\text{v}}}\sqrt{\pi}D_{\text{c}}}\right)}^2
 \label{carrier estimation diffusion}
\end{equation}

For our device, we measure $R_{\text{sq}} \sim 400~\Omega$. In the absence of the magnetic moments, the charge ($D_{\text{c}}$) and the spin spin diffusion coefficient ($D_{\text{s}}$) will be equal\cite{94_Weber2005_Nat_DcDs}. From the spin transport measurements, we extract $D_{\text{s}}=0.04~m^2/s$ and use this value to estimate $n$ in the graphene flake from Eq.~\ref{carrier estimation diffusion} $\sim$ 5*10$^{12}$ cm$^{-2}$. Using the relation $\sigma= ne\mu$, we estimate the carrier mobility $\mu \sim$ 3000 cm$^2$V$^{-1}$s$^{-1}$.

When a bias is applied across a cobalt/2L-hBN tunnel barrier, it modifies the carrier density underneath the contact \cite{bokdam_electrostatic_2011}. In order to estimate this, we assume that initially, the graphene is undoped ($E_{\text{F}}=0$) underneath the contact. However, the actual doping is unknown. On applying the bias $V$, the Fermi level is changed by $\triangle E_{\text{F}}$:
\begin{equation}
\triangle n=\frac{g_{\text{s}}g_{\text{v}}\pi \triangle E_{\text{F}}^2}{h^2 v_{\text{F}}^2}
 \label{fermi}
\end{equation}
 
which can be related with the external bias $V$ with the following relation:
\begin{equation}
\triangle n=C_{\text{o}}(V-\frac{\triangle E_{\text{F}}}{e})=\frac{\epsilon_0 \epsilon_r}{d}(V-\frac{\triangle E_{\text{F}}}{e})
 \label{carrier density}
\end{equation}
Here, $C_{\text{o}}$ is the geometrical capacitance of the 2L-hBN tunnel barrier, $\epsilon_0$ is the dielectric permittivity ($=8.85 \times 10^{-12}$ F/m), $\epsilon_r$ is the relative dielectric permittivity of the hBN ($\sim$ 4), $e$  is the electronic charge, and $d$ is the thickness of the tunnel barrier (= $\SI{7}{\angstrom}$). Now, we can obtain $\triangle E_{\text{F}}$ by combining Eq.~\ref{fermi} and \ref{carrier density}:

\begin{equation}
 \triangle E_{\text{F}}=\frac{\pm \sqrt{1+4ceV}-1}{2c}
 \label{n_cq}
\end{equation}

where $c=(4\pi de^2)/(h^2v_{\text{F}}^2\epsilon_0 \epsilon_r)$

We obtain $\triangle E_{\text{F}}$ and $\triangle n$ from the equations~\ref{n_cq} and \ref{fermi}. For the applied bias $V\sim$ $\pm$ 0.6 V across the tunnel barrier, $n$ can be modified up to $\pm$ 8*10$^{12}$ cm$^{-2}$, implying that it is possible to tune the carrier density underneath the contact from p- to n-type or vice versa around the charge neutrality point.   







\section{Drift effects on spin injection/detection polarization and spin transport}
Jozsa et al.\cite{31_Jozsa2009_PRB} reported an enhanced differential spin-injection polarization using the pinhole Al$_{\text{2}}$O$_{\text{3}}$ barriers from 18$\%$ at zero DC current bias upto 31$\%$ at +5 $\mu$A bias, while it approaches zero at reverse bias due to a strong local carrier drift near the low resistive regions beneath the contact. On the contrary, we observe an increase in the magnitude of the differential polarization and a change in the sign on reversing the bias. 
This indicates that the observed behaviour in our device is not due to the carrier drift.

The presence of a non-zero electric-field in the graphene spin transport channel could also modify $\lambda_{\text{s}}$. The spin relaxation length due to the positive drift field (upstream of spins)$\lambda_{+}$, and due to the negative  drift field (downstream of spins) $\lambda_{-}$ can be calculated from\cite{79_Ingla-Aynes2016_NL}
\begin{equation}
\frac{1}{\lambda_{\pm}}=\pm \frac{v_{\text{d}}}{2D_{\text{s}}} + \sqrt{{\left(\frac{1}{\sqrt{\tau_{\text{s}}D_{\text{s}}}}\right)}^2+{\left(\frac{v_{\text{d}}}{2D_{\text{s}}}\right)}^2} 
\end{equation}
Here $v_{\text{d}}=\mu E$ is the drift velocity of the electron(or hole) in an electric-field $E = IR_{\text{sq}}/L $, $\mu$ is the field-effect carrier mobility, and L is length of the spin-transport channel.  For an applied bias of 20 $\mu$A and channel length of 1 $\mu$m with a carrier mobility $\sim$ 3000 cm$^2$V$^{-1}$s$^{-1}$, the calculations lead to $\lambda_{+}$ = 4.9 $\mu$m and $\lambda_{-}$ = 6.7 $\mu$m, whereas the spin relaxation length obtained from the Hanle fitting, under zero bias, is 5.8 $\mu$m which is nearly equal to the average of $\lambda_{+}$ and $\lambda_{-}$. The polarization values, obtained using $\lambda_{+}$ or $\lambda_{-}$, differ by 10\%, compared to that extracted using $\lambda_{\text{s}}$ in the absence of the drift field. This implies that the injector and detector polarizations also have a similar uncertainty.


\section{Bias dependence for Co/TiO$_{\text{2}}$/graphene tunneling contacts}
\begin{figure}[!htbp]
\centering
 \includegraphics[width=\columnwidth,trim= 0in 0in 0in 0in,clip]{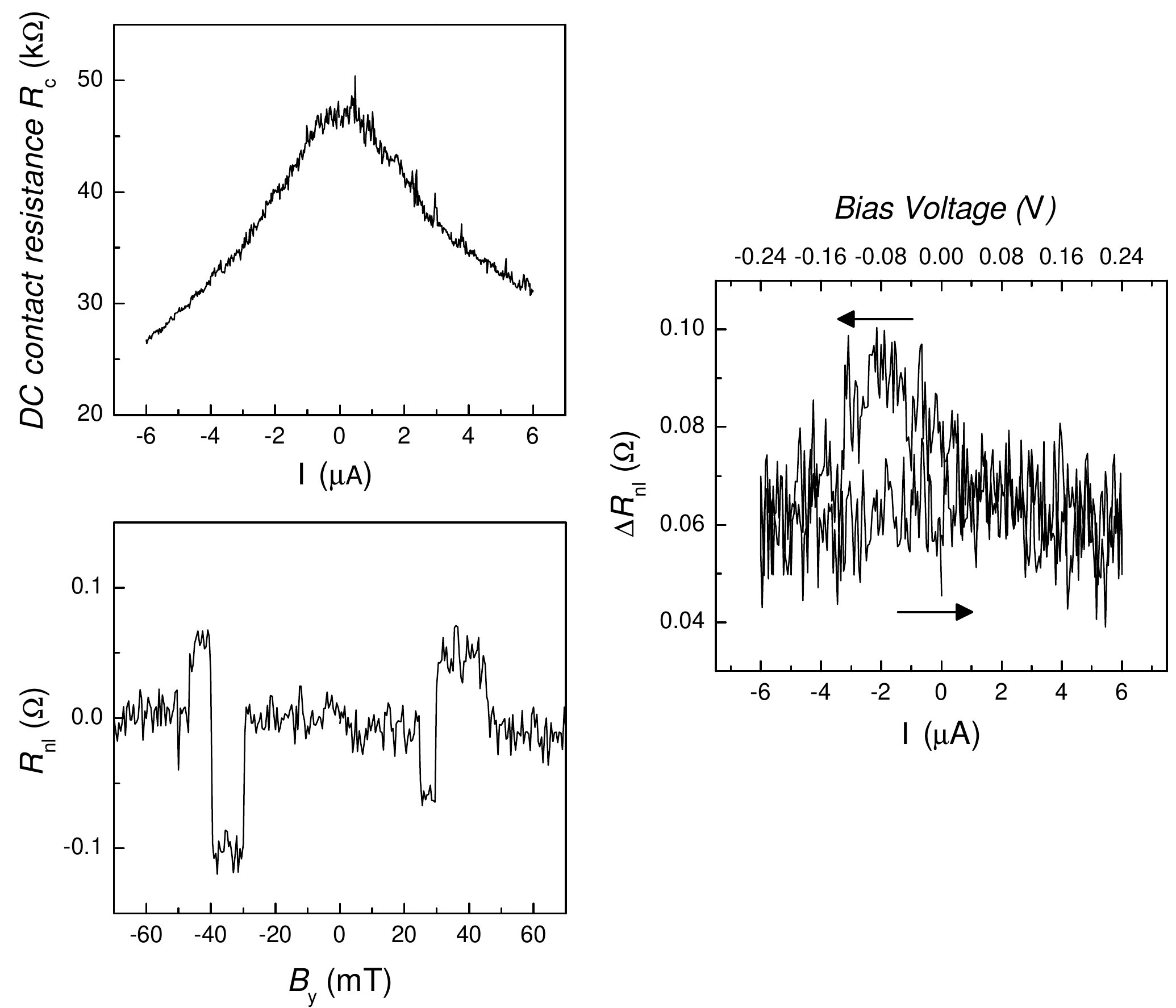}
 \caption{(\textbf{a}) DC contact resistance $R_c(=V/I$) of a Co/TiO$_{\text{2}}$/graphene tunnel barrier shows a non-linear behaviour as a function of DC current bias $I$, implying a tunneling behaviour of contacts.  (\textbf{b}) A spin-valve measurement for graphene with TiO$_{2}$ tunnel barriers. An offset at zero field is subtracted from the non-local resistance. (\textbf{c}) Non-local spin signal $\triangle R_{\text{nl}}$ for the spin injection through an injector electrode with TiO$_{\text{2}}$ tunnel barrier a function of DC current bias. Arrows indicate the direction of the bias sweep. In contrast to the contacts with 2L-hBN tunnel barriers (Fig.~\ref{Low_Rc_1L_2LhBN}), the contacts with TiO$_{2}$ barriers show no change in the magnitude and the sign of the injection polarization as a function of $I$.}
 \label{TiO2_polarization}
\end{figure}

We also perform the same experiment on a reference sample with TiO$_{\text{2}}$ tunnel barriers. The contact resistance for FM electrodes with the TiO$_{\text{2}}$ was around 40 k$\Omega$ which is comparable to the interface resistance of the contacts with a 2L-hBN tunnel barrier. However, we do not see any sign reversal of the non-local spin-signal ($\triangle R_{\text{nl}}$) within the range of applied bias $I$ on injector contact. Also, the magnitude of $\triangle R_{\text{nl}}$ is hardly modified (Fig.~\ref{TiO2_polarization}).

\newpage






\end{document}